\documentclass[sn-mathphys-num,pdflatex]{sn-jnl}% Math and Physical Sciences Numbered Reference Style 
\usepackage{graphicx}%
\usepackage{multirow}%
\usepackage{amsmath,amssymb,amsfonts}%
\usepackage{amsthm}%
\usepackage{mathrsfs}%
\usepackage[title]{appendix}%
\usepackage{xcolor}%
\usepackage{textcomp}%
\usepackage{manyfoot}%
\usepackage{booktabs}%
\usepackage{algorithm}%
\usepackage{algorithmicx}%
\usepackage{algpseudocode}%
\usepackage{listings}%
\usepackage{bm}
\usepackage{geometry}
\begin{document}

\title[Domains in ferroelectric nematic liquid crystals]{Twist, splay, and uniform domains in ferroelectric nematic liquid crystals}

\author*[1,2]{\fnm{Maxim O.} \sur{Lavrentovich}}\email{lavrentm@gmail.com}
\equalcont{These authors contributed equally to this work.}

\author[3,4]{\fnm{Priyanka} \sur{Kumari}} 
\equalcont{These authors contributed equally to this work.}

\author*[3,4,5]{\fnm{Oleg D.} \sur{Lavrentovich}}\email{olavrent@kent.edu} 

\affil[1]{\orgdiv{Department of Earth, Environment, and Physics}, \orgname{Worcester State University}, \orgaddress{\city{Worcester}, \state{MA} \postcode{01602}, \country{USA}}}

\affil[2]{\orgdiv{Department of Physics \& Astronomy}, \orgname{University of Tennessee}, \orgaddress{\city{Knoxville}, \state{TN} \postcode{37996}, \country{USA}}}

\affil[3]{\orgdiv{Advanced Materials and Liquid Crystal Institute}, \orgname{Kent State University}, \orgaddress{\city{Kent}, \state{OH} \postcode{44242}, \country{USA}}}

\affil[4]{\orgdiv{Materials Science Graduate Program}, \orgname{Kent State University}, \orgaddress{\city{Kent},  \state{OH} \postcode{44242},  \country{USA}}}

\affil[5]{\orgdiv{Department of Physics}, \orgname{Kent State University}, \orgaddress{\city{Kent},  \state{OH} \postcode{44242}, \country{USA}}}


\abstract{The newly-discovered ferroelectric nematic liquid crystal exhibits a variety of unique defect phenomena. The depolarization field in the material favors spontaneous spatial variations in polarization, manifesting in diverse forms such as bulk twists and arrangements of alternating polarization domains. The configuration of these domains is governed by a balance between depolarization field reduction and molecular alignment at interfaces. We investigate a ferroelectric nematic confined in a thin cell with apolar surface anchoring, patterned using photoalignment. Under uniform planar alignment, the system forms stripes, while a radial +1 defect pattern results in pie-slice domains. Neighboring domains show either opposite directions of uniform polarization (thin cells) or opposite handedness of the spontaneous twist (thick cells). Our calculations and experiments demonstrate that electrostatic interactions tend to shrink domain size, whereas elastic and surface anchoring effects promote larger domains. In this work, we make predictions and measurements of the domain size as a function of cell thickness, and show that ionic screening suppresses domain formation.}


\maketitle

\newpage

\section*{Introduction}

The polar ordering of the recently discovered ferroelectric nematic ($\mathrm{N}_\mathrm{F}$) liquid crystal \cite{mandle2017nematic,nishikawa2017fluid,sebastian2020ferroelectric,chen2020first} creates a fascinating interplay between the elasticity, surface interactions, and the electrostatic energy associated with spatial variations of the spontaneous polarization $\mathbf{P}$. The ensuing $\mathrm{N}_\mathrm{F}$ structures are not restricted by crystallographic axes and can be studied by polarizing optical microscopy since, in the materials explored so far, the polarization $\mathbf{P}$ is along the optic axis, or the director $\hat{\mathbf{n}}$.

$\mathrm{N}_\mathrm{F}$ structures tend to avoid a splay deformation (divergence of $\mathbf{P}$) since splay creates a bound charge of a bulk density $\rho = - \nabla \cdot \mathbf{P}$ and increases the electrostatic energy. Polarization-related charges are also avoided at surfaces and at domain walls. For example, a uniform polarization, $\mathbf{P}(x,y,z)=\text{const.}$, would deposit charges at the opposite ends of the sample and create a strong, energetically costly depolarization field: $E=-P/\epsilon \epsilon_0 \approx 10^{8}~\mathrm{V}/\mathrm{m}$; here $\epsilon \approx (10-100)$ \cite{adaka2024dielectric,erkoreka2024dielectric} and $P \approx (3-7) \times 10^{-2}~\mathrm{C}/\mathrm{m}^2$ are the typical permittivity and polarization of the $\mathrm{N}_\mathrm{F}$ phase. Polydomain textures of thin $\mathrm{N}_\mathrm{F}$ films that impose no preferred in-plane orientation of $\mathbf{P}$, such as a film supported by an isotropic fluid \cite{solitons2022,FNconics2023} or freely suspended in air \cite{clark2024freely}, present clear evidence of these tendencies. First, the spontaneous polarization is everywhere in the plane of the film, avoiding surface charges, which would occur whenever $\mathbf{P}$ is tilted. Second, the in-plane textures are dominated by two types of domains, in which $\mathbf{P}$ is either uniform or bends into circular vortices, thus avoiding splay deformations and associated space charge in the bulk. Domain walls separating these domains adopt the shapes of conic sections, such as parabolas and hyperbolas \cite{solitons2022,FNconics2023,clark2024freely}.

The $\mathrm{N}_\mathrm{F}$ structure changes dramatically when one of the film's surfaces imposes a unidirectional alignment of $\mathbf{P}$ and the other is azimuthally degenerate. In this case, one might expect a uniform state since circular vortices are not compatible with the unidirectional surface alignment. However, experiments \cite{sciencetwist2024} demonstrate that instead of being uniform, the polarization twists around the   film normal, so that the vectors $\mathbf{P}$ are antiparallel to each other  at the bottom and the top surfaces, thus mitigating the depolarization effect. The twisted structures arise from the balance of electrostatic and elastic energies. Clearly, this balance should be affected by the geometry of confinement (e.g., a film with a large lateral extension or a long cylinder), by surface anchoring at the bounding plates, and by the free ions capable of at least partial screening of   bound charges.

 In this work, we explore experimentally and theoretically the interplay between electrostatic and elastic energies in  $\mathrm{N}_\mathrm{F}$ domain structures, taking into account the effects of surface anchoring and ionic screening. The surface anchoring in the experiments is designed to be in-plane apolar (bidirectional) by using a photoalignment technique \cite{guo2016high,guo2016designs}. There are two reasons: First, apolar anchoring should avoid twists that are artificially created by the antiparallel assembly of two plates with a unidirectional alignment of $\mathbf{P}$, which happens for mechanically rubbed plates. Second, the photoalignment technique  allows one to impose various patterns of surface alignment of $\mathbf{P}$ (including an unwelcome splay), thereby exploring the electrostatics-elasticity balance in different contexts.

We find that, generally, the electrostatic interaction prefers a spatial modulation of $\mathbf{P}$ with a characteristic size $\lambda$ (e.g., the twist pitch or the domain size), generating an energetic contribution that increases with $\lambda$. Conversely, any such modulation incurs an elastic or anchoring energy penalty, which  decreases with $\lambda$. Thus, the balance between elastic (or anchoring) energy and the electrostatic interaction generates a preferred value of $\lambda$, leading to the various patterns considered here.

 We   present experimental studies of domain structures in flat slabs with bidirectional anchoring $   \hat{\mathbf{n}}_0$ designed to set uniform planar or radial patterns of molecular orientations at the bounding surfaces.  We observe that the $\mathrm{N}_\mathrm{F}$ avoids monocrystal alignment of $\mathbf{P}$ by forming $\pi$-twisted striped domains ($\pi$-TDs) in planar cells, provided they are sufficiently thick (a few microns). The twist axis is perpendicular to the bounding plates. In thin (micron or less) planar cells, the electrostatic energy is reduced by forming a periodic lattice of elongated  uniform domains (UDs) with constant $\mathbf{P}$ parallel to $\hat{\mathbf{n}}_0$ but alternating the polarity from one domain to the next. In the radial splay patterns, the surface-imposed divergence of $\mathbf{P}$ is relaxed by the TDs in thick cells and by  pie slices of splay domains (SDs) in thin cells, with the polarization alternatively pointing toward or away from the $+1$ defect core.  We develop theories for both the stripe width and the number of SD slices.  We compare our theoretical predictions and experimental measurements of the number of domain walls (SD slices or stripe widths) in cells of varying thickness $h$.

\section*{Results}

\subsection*{Stripe patterns in planar cells}

\begin{figure}[ht]
\centering
\includegraphics[width=\textwidth]{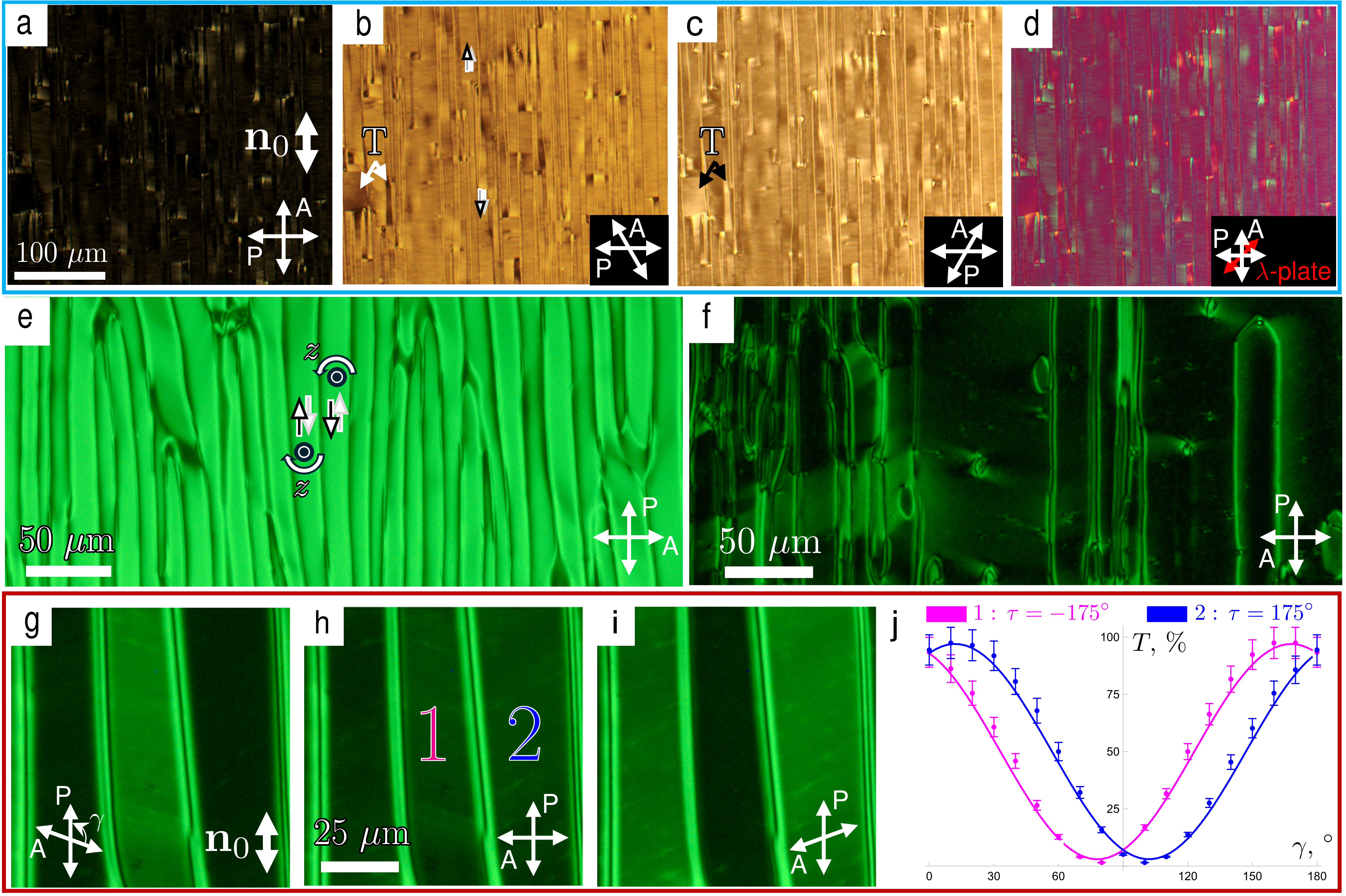}
\caption{\label{fig:Planar}  \textbf{Domain structures in photoaligned cells with planar apolar anchoring}. \textbf{a} A polarizing optical microscopy texture of uniform domains (UDs) in a thin $\mathrm{N}_\mathrm{F}$ cell, $h= 0.7~\mu\mathrm{m}$ (temperature $T=55^{\circ}\mathrm{C}$) with bidirectional anchoring indicated by the white double arrow. \textbf{b},\textbf{c} The same, counterclockwise and clockwise uncrossing of analyzer and polarizer, respectively. Pairs of parallel arrows in (\textbf{b}) illustrate that $\mathbf{P}$ does not change along the $z$-axis normal to the cell but alternates from one UD to the next along the $x$-axis. \textbf{d} The same, observation with an optical compensator; the slow axis is along the red double arrow.  Most domains are uniform in (\textbf{a}-\textbf{d}), with the exception of small areas labeled with a letter T, indicating a $\pi$-twisted region. \textbf{e} Polarizing optical microscopy texture of a thick $\mathrm{N}_\mathrm{F}$ cell, $h= 5~\mu\mathrm{m}$ ($T=60^{\circ}\mathrm{C}$). Light transmission indicates that the sample is split into $\pi$-twisted domains ($\pi$-TDs), shown schematically by two antiparallel arrows that twist by $\pi$ around the $z$-axis. \textbf{f} Domains are suppressed when DIO is doped with 0.5 wt.\% of an ionic fluid BMIM-$\mathrm{PF}_\mathrm{6}$, shown for a cell with $h=4.0~\mu\mathrm{m}$ ($T=43^{\circ}\mathrm{C}$).  \textbf{g},\textbf{h},\textbf{i}  The twisted polarization in the $\pi$-TDs is readily recognized by observing the textures with uncrossed (\textbf{g},\textbf{i}) and crossed (\textbf{h}) polarizers, for a cell with $h=4.0~\mu\mathrm{m}$ ($T=60^{\circ}\mathrm{C}$). Textures (\textbf{e}-\textbf{i}) are captured using a green interferometric filter with a center wavelength $\lambda=532~\mathrm{nm}$ and    $1~\mathrm{nm }$    bandwidth. \textbf{j} Theoretical fits (lines) to light transmission data (points) measured through small localized regions in the two domains highlighted in (\textbf{h})  as a function of the angle $\gamma$ between the polarizers. Fits yield twist angles $\tau =\pm 175^{\circ}$ (see Methods) in the two adjacent domains. Error bars are  standard deviations of the intensity fluctuations. } 
\end{figure}

The planar cells in the N and $\mathrm{SmZ}_\mathrm{A}$ phases show homogeneous textures with the molecular director $\hat{\mathbf{n}}$ parallel to the photoinduced apolar easy axis  $ \pm\hat{\mathbf{n}}_0=(0,\pm{1},0)$. Here and in what follows, we use the Cartesian coordinates $(x,y,z)$ in which the $y$-axis is the direction of the   anchoring and the $z$-axis is normal to the film.
Upon cooling, the texture remain homogeneous for about $(4-8)~{}^{\circ}\mathrm{C}$ below the $\mathrm{SmZ}_\mathrm{A}$-$\mathrm{N}_\mathrm{F}$ transition point, depending on the cell thickness. Thin cells, $h< 2~\mu\mathrm{m}$, preserve uniformity, $\mathbf{P}=P(0,\pm{1},0)$ for $(6-8)^{\circ}\mathrm{C}$, after which they split into a lattice of UDs elongated along $\hat{\mathbf{n}}_0$, each of width on the order of $10~\mu\mathrm{m}$, Fig.~\ref{fig:Planar}a-d. When $\hat{\mathbf{n}}_0$ is parallel to one of the polarizers, the UDs are practically extinct between two crossed polarizers, Fig.~\ref{fig:Planar}a, and their optical retardance equals that of the optical compensator when the latter is inserted between the sample and the analyzer, Fig.~\ref{fig:Planar}d. The textures observed with polarizers uncrossed counterclockwise, Fig.~\ref{fig:Planar}b, and clockwise, Fig.~\ref{fig:Planar}c, differ little from each other. One concludes that $\mathbf{P}$ in UDs aligns along $\hat{\mathbf{n}}_0$ and their polarity alternates from $\mathbf{P}=P_0(0,1,0)$ in one domain to $\mathbf{P}=P_0(0,-1,0)$ in the next. There are only few regions, marked with a letter T in Figs.~\ref{fig:Planar}b,c, in which the textures with uncrossed polarizers do differ, which suggests a twist of $\mathbf{P}$ along the $z$-axis.

Thick cells, $h> 2~\mu\mathrm{m}$, show a very different behavior. Below $62^{\circ}\mathrm{C}$, they develop a stripe pattern of $\pi$-twisted domains ($\pi$-TDs), recognized by the absence of light extinction when viewed between two crossed polarizers, one of which is along $\hat{\mathbf{n}}_0$, as seen in Fig.~\ref{fig:Planar}e. This $\mathrm{N}_\mathrm{F}$ texture is similar to the previously studied TDs with alternating left-handed and right-handed twists in cells in which one plate sets a unipolar alignment of $\mathbf{P}$ and the other is azimuthally degenerate \cite{sciencetwist2024}.\par
The addition of an ionic salt 1-Butyl-3-methylimidazolium hexafluorophosphate (BMIM-$\mathrm{PF}_\mathrm{6}$) suppresses the $\pi$-TDs, as the sample is predominantly extinct between the crossed polarizers, Fig.~\ref{fig:Planar}f. The added salt also decreases the temperatures of phase transition by approximately $  20{}^{\circ}\mathrm{C}$, in agreement with previous studies \cite{zhong2023thermotropic,sciencetwist2024,antiFN2025}.  The dependency of a transmitted light intensity on the angle $\gamma$ between the directions of polarization of the polarizer and analyzer allows one to determine that the twist angle $\tau$ between the bottom and top orientation of $\mathbf{P}$ in the DIO cell of a thickness $h=4~\mathrm{\mu m}$ is close to $180^{\circ}$, Fig.~\ref{fig:Planar}g-j.

Note that the domain structures presented in Fig.~\ref{fig:Planar} are different from the recently described splay and double-splay domain structures that form in the material RM734 above the phase transition to the $\mathrm{N}_{\mathrm{F}}$ \cite{antiFN2025,weidoublesplay2024}. These splay domain structures are optically discernable when the material is exposed to an ionic fluid \cite{antiFN2025} or an ionic polymer \cite{weidoublesplay2024}. The splay domains above the $\mathrm{N}_{\mathrm{F}}$ temperature range are attributed to the flexoelectric effect, which favors the splay of molecules with head-tail asymmetry \cite{antiFN2025,weidoublesplay2024}. In the $\mathrm{N}_{\mathrm{F}}$  phase, this flexoelectricity-triggered splay is suppressed by the space charge that accompanies divergence of $\mathbf{P}$ \cite{antiFN2025}. As a result, the domain structures in the  $\mathrm{N}_{\mathrm{F}}$ phase, Fig.~\ref{fig:Planar}, are different from the domain structures reported in Refs. \cite{antiFN2025,weidoublesplay2024}. Thus, in what follows, we disregard the flexoelectric effect.

\subsection*{Pie slices in cells pre-patterned with a $+1$ radial splay defect}

\begin{figure*}[htp]
\centering
\includegraphics[width=\textwidth]{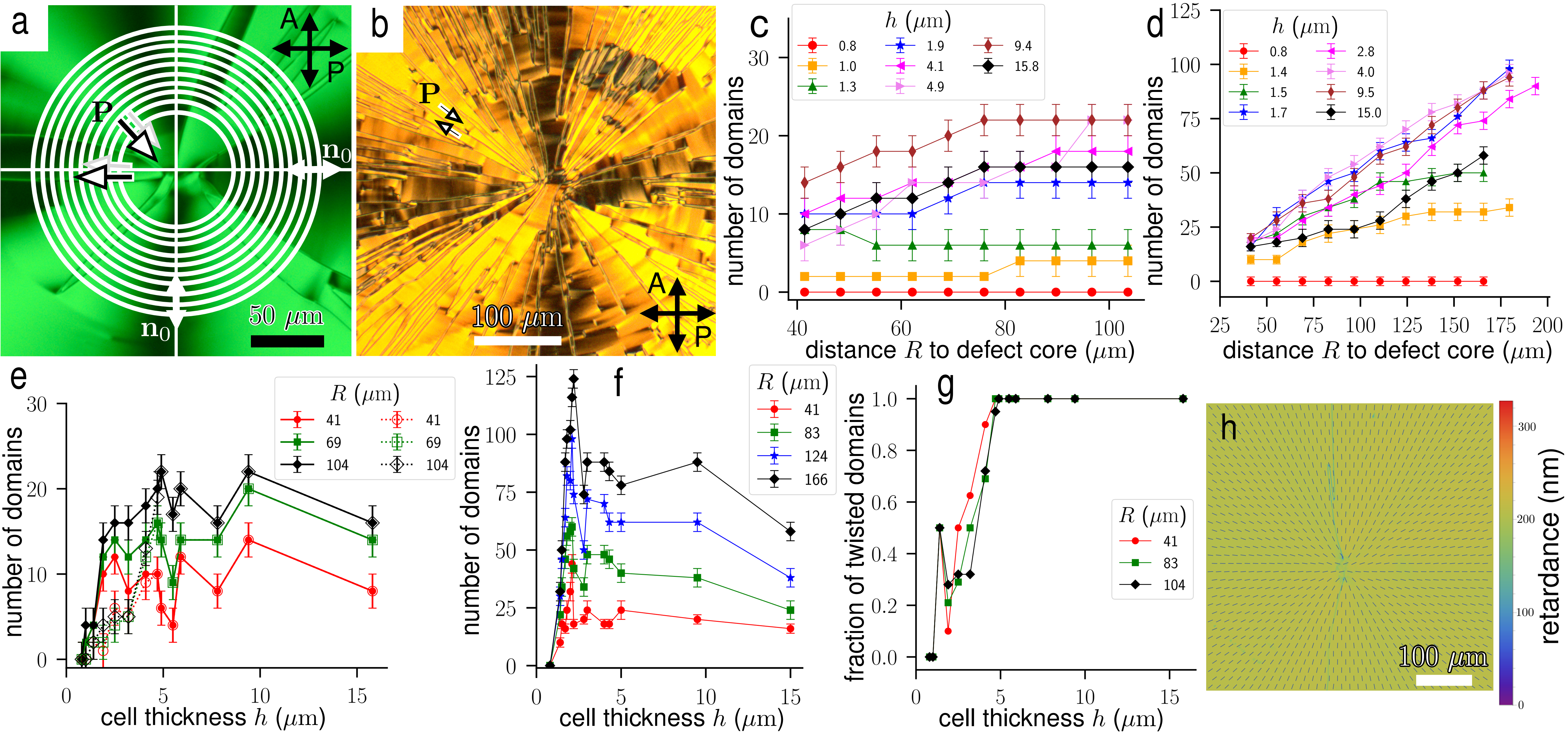}
\caption{\label{fig:SD} \textbf{Characteristics of a +1 radial pre-patterned splay defect.}  \textbf{a} A polarizing optical microscopy texture (recorded in a monochromatic light using a green filter with $532~\mathrm{nm}$  wavelength and    $1~\mathrm{nm}$ bandwidth) of an $\mathrm{N}_\mathrm{F}$ cell with thickness $h= 2~\mu\mathrm{m}$, $T=55^{\circ}\mathrm{C}$, and cooling rate $0.1^{\circ}\mathrm{C}$ /min. The parallel arrows show  the parallel polarizations at the bottom and top plates, with $\pi$ flips from one splay domain (SD) to the next along the azimuthal direction. \textbf{b} Similar cell but recorded under white light,  $h= 2~\mu\mathrm{m}$, $T=55^{\circ}\mathrm{C}$, and cooling rate $5^{\circ}\mathrm{C}$/min. \textbf{c},\textbf{d} Number of domains as a function of distance $R$ from the defect core for cell thicknesses $h= (1-16)~\mu\mathrm{m}$, cooling rates $0.1^{\circ}\mathrm{C}$ and $5^{\circ}\mathrm{C}$, respectively. \textbf{e} Number of domains versus $h$ measured at different distances $R$ from the defect core, cooling rate $0.1^{\circ}\mathrm{C}$/min. Filled symbols and solid lines correspond to the total domain count, while open symbols and dashed lines correspond to the number of  $\pi$-TDs. \textbf{f} The $h$-dependence of the number of domains, cooling rate $5^{\circ}\mathrm{C}$/min. Error bars in (\textbf{c}-\textbf{f}) are estimates of uncertainties in the domain count in the sample. \textbf{g} Fraction of TDs as a function of the cell thickness $h$ measured at different distances $R$ from the defect core; cooling rate $0.1^{\circ}\mathrm{C}$ /min.  \textbf{h} Radial structure with no domains in a thin cell, $h=1~\mu\mathrm{m}$ ($T=50{}^{\circ}\mathrm{C}$), imaged using the Microimager PolScope with the ticks showing the director field $\hat{\mathbf{n}}(x,y)$. The thin blue line of decreased retardance is a domain wall separating the  $+1/2$ defects that split from the central $+1$ splay pattern.} 
\end{figure*}

\begin{figure}[htp]
\centering
\includegraphics[width=0.4\textwidth]{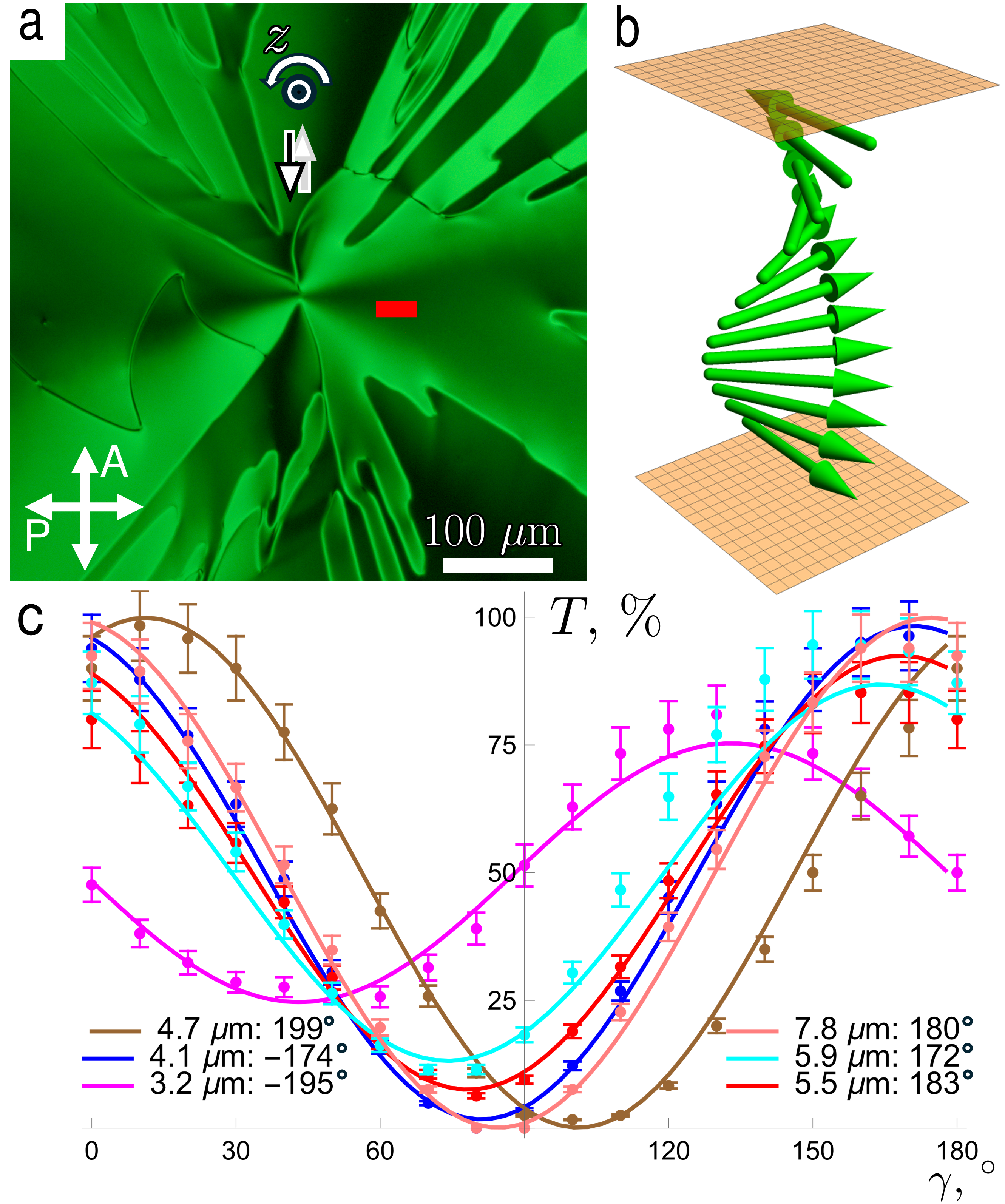}
\caption{\label{fig:TD}  \textbf{Twist in a +1 radial defect pre-patterned cell.} \textbf{a} A polarizing optical microscopy texture (recorded in a monochromatic light using a green filter with $532~\mathrm{nm}$  wavelength and    $1~\mathrm{nm}$ bandwidth) of an $\mathrm{N}_\mathrm{F}$ in a cell of the thickness $h= 5.5~\mu\mathrm{m}$ (DIO, temperature $T=60^{\circ}\mathrm{C}$). The curved white arrow shows the direction of polarization twist and the anti-parallel arrows indicate the $\pi$-twist of $\mathbf{P}$ between the top and bottom plates.  \textbf{b} Schematic of the polarization $\mathbf{P}$ (green arrow) performing a $\pi$-twist between the bottom and top cell surfaces (orange). \textbf{c} Fitting the dependence of transmitted light intensity [through a small region in a domain like the red rectangle in \textbf{a}] on the angle $\gamma$ between the polarizers yields the twist angle $\tau \approx 180^{\circ}$ for samples of different thickness in the range $3~\mu\mathrm{m}<h<8 ~\mu\mathrm{m}$, demonstrating the $\pi$-twist (see Methods). The error bars are the standard deviations of light intensity fluctuations in the measured small regions.}
\end{figure}

The radial patterns of the +1 defect in the N and $\mathrm{SmZ}_\mathrm{A}$ phases demonstrate smooth splay deformation of the director. In the $\mathrm{N}_\mathrm{F}$ phase, the textures show domains of two types: The first are  pie slices, or splay domains (SDs), with $\mathbf{P}$ parallel or anti-parallel to the  radial direction $\hat{\mathbf{r}}$,  pointing either away from the core of the +1 defect at $\hat{\mathbf{r}}=0$ or towards it, as shown in Fig.~\ref{fig:SD}a,b. As established by polarizing microscopy, within each SD, $\mathbf{P}$  does not twist along the $z$ axis, except perhaps within the domain walls that separate   sectors of antiparallel $\mathbf{P}$, similar to the UDs in the planar cells. The second type are $\pi$-TDs, similar to the $\pi$-TDs in the planar cells: $\mathbf{P}$ twists around the $z$ axis by $\tau \approx 180^{\circ}$, as shown in Fig.~\ref{fig:TD}.

 Thick cells, $h>6~\mathrm{\mu m}$,  upon cooling below $61^{\circ}\mathrm{C}$, show exclusively $\pi$-TDs, Fig.~\ref{fig:SD}e,g. Cells of an intermediate thickness, $1~\mathrm{\mu m}<h<6~\mathrm{\mu m}$, show both $\pi$-TDs and SDs, Fig.~\ref{fig:SD}e,g. Structures in cells with $h \le 1~\mathrm{\mu m}$ are either uniformly radial, Fig.~\ref{fig:SD}h, or show a few SDs. In all of these cases, there is splay deformation originating from the $+1$ central vortex and the SDs and $\pi$-TDs terminate in a complex manner near this central defect region (i.e., at the tips of the pie slices), as seen in Fig.~\ref{fig:SD}b.   Thus, when comparing these samples to our theoretical results below, we will focus on the region away from this complex region, out at distances $R \approx 100~\mu\mathrm{m}$ from the defect center where the sectors do not terminate but where we still find substantial splay deformation and, consequently, the presence of bound charge.

  \begin{figure}[htp]
\centering
\includegraphics[width=0.4\textwidth]{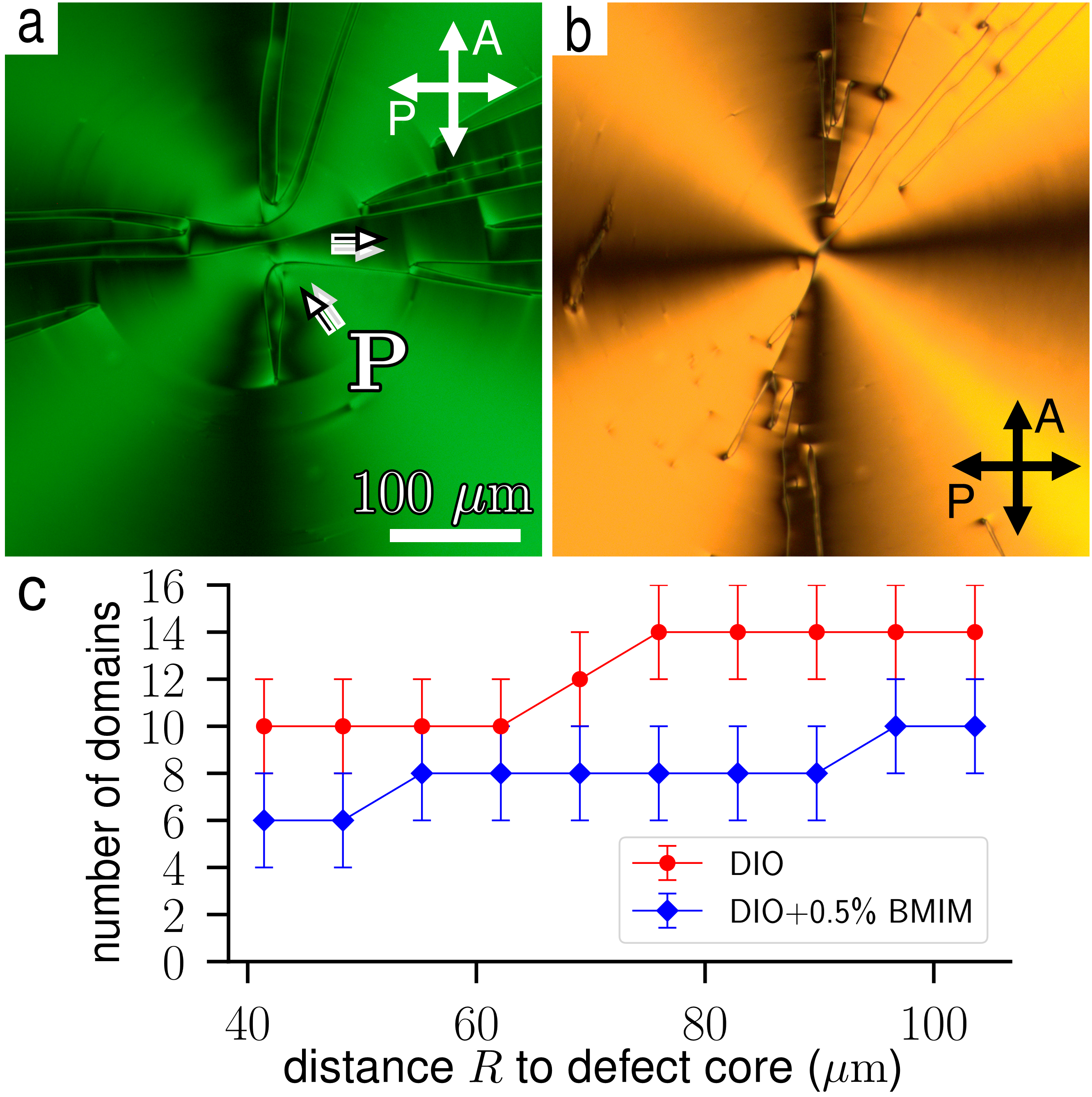}
\caption{\label{fig:BMIM} \textbf{Effects of ion addition.}  A polarizing optical microscopy texture (recorded in a monochromatic light using a green filter with $532~\mathrm{nm}$  wavelength and    $1~\mathrm{nm}$ bandwidth)  of the $\mathrm{N}_\mathrm{F}$ phase of, \textbf{a}, pure DIO, $T=50^{\circ}\mathrm{C}$ and \textbf{b}, DIO doped with 0.5\% BMIM-$\mathrm{PF}_\mathrm{6}$, $T=43^{\circ}\mathrm{C}$ in cells with radial patterns of the thickness $h= 1.9~\mu\mathrm{m}$. \textbf{c} Number of domains versus distance to the +1 defect center for pure DIO and DIO doped with BMIM-$\mathrm{PF}_\mathrm{6}$. Error bars indicate that it was difficult to distinguish about two of the sectors in the two samples.} 
\end{figure}

In thin cells, $1~\mathrm{\mu m}<h<6~\mathrm{\mu m}$, the domains form at about $(5-10)^{\circ}\mathrm{C}$ below the $\mathrm{SmZ}_\mathrm{A}$-$\mathrm{N}_\mathrm{F}$ transition point. The thickness dependency of the temperature at which the domains appear can be qualitatively explained by the temperature dependence of the polarization magnitude $P_0$ which increases from about $P_0=3.4 \times 10^{-2}~\mathrm{C}/\mathrm{m}^2$ to $P_0=4.6 \times 10^{-2}~\mathrm{C}/\mathrm{m}^2$ \cite{nishikawa2017fluid} as the temperature decreases following the $\mathrm{SmZ}_\mathrm{A}$-$\mathrm{N}_\mathrm{F}$ transition; the bound charge effects leading to the domains might be weaker than the stabilizing surface anchoring and ion screening.

The addition of either a $\pi$-TD or SD  domain increases the effective charge of the central  defect as the polarization orientation will twist by an additional factor of $2\pi$ for each pie slice in the domain configuration. The central $+1$ defect in the pattern, then, breaks up into a complex arrangement where all of the pie slice corners merge together. When there are no sectors, such as in thin cells with $h \sim 1~\mu\mathrm{m}$,   the $+1$ defect splits into two $+1/2$ defect cores separated by a domain wall, as illustrated in Fig.~\ref{fig:SD}h with a thin blue line of a somewhat smaller retardance ($\sim 160~\mathrm{nm}$   versus $\sim 210~\mathrm{nm}$ in the far-field).  This result indicates that there is no melting of the ferroelectric order (in which case the domain wall will be absent) nor significant escape into the third dimension (in which case the central part would  have significantly lower retardance).  In a regular nematic $\mathrm{N}$, the $+1$ defect in the center would break up into disclination lines or contain an escaped structure \cite{nematicplusone}. The absence of the escape into the third dimension in the $\mathrm{N}_{\mathrm{F}}$ is also evident in other textures in Figs. \ref{fig:SD}, \ref{fig:TD} and \ref{fig:BMIM}. This is not surprising since, in an $\mathrm{N}_\mathrm{F}$, such an escape of $\mathbf{P}$ along the normal to the cell would either deposit surface charges at the bounding plates or create additional splay with an accompanying bulk space charge.  Note that even in a paraelectric N, a $+1$ defect prefers to split into a pair of $+1/2$ defects when the cell is sufficiently thin (see \cite{nematicplusone,nematicplusonePRE,photopatternanchoring}).

Within the range of coexistence, the fraction of the SD domains increases as the cell thickness $h$ decreases, Fig.~\ref{fig:SD}e,g. The number of domains increases with the cooling rate, Fig.~\ref{fig:SD}c,d.
One potential reason is that for a longer cooling time, the highly polar material absorbs ions from the surroundings, such as glue, BY layer, etc., to screen the bound charge. The ion concentration of free ions in the $\mathrm{N}_\mathrm{F}$ cannot be measured directly by conventional techniques since   polarization reorientation reduces the electric field in the $\mathrm{N}_\mathrm{F}$ bulk \cite{clark2024dielectric}. Briefly heating the sample to $120^{\circ}\mathrm{C}$ into the N phase and measuring the concentration supports the idea that ion concentration can increase in time: We find $c(0) = 5.0 \times 10^{22}~\mathrm{ions}/\mathrm{m}^{3}$ at the start of experiment to $c(18~\mathrm{hours}) = 6.3 \times 10^{22}~\mathrm{ions}/\mathrm{m}^{3}$ after 18 hours of keeping the sample in the $\mathrm{N}_\mathrm{F}$ phase at $65^{\circ}\mathrm{C}$.
The number of domains decreases significantly when DIO is doped with the ionic fluid BMIM-$\mathrm{PF}_\mathrm{6}$, Fig.~\ref{fig:BMIM}. At a weight concentration 0.5 wt \% of BMIM-$\mathrm{PF}_\mathrm{6}$, (an ion concentration of $1.5 \times 10^{25}~\mathrm{ions}/\mathrm{m}^{3}$ for fully ionized added molecules), the $\mathrm{N}_\mathrm{F}$ phase preserves ferroelectric ordering, as reported by Zhong et al. \cite{zhong2023thermotropic}.

\subsection*{Theory of twisted cylindrical domains}

\begin{figure}[htp]
\centering
\includegraphics[width=0.4\textwidth]{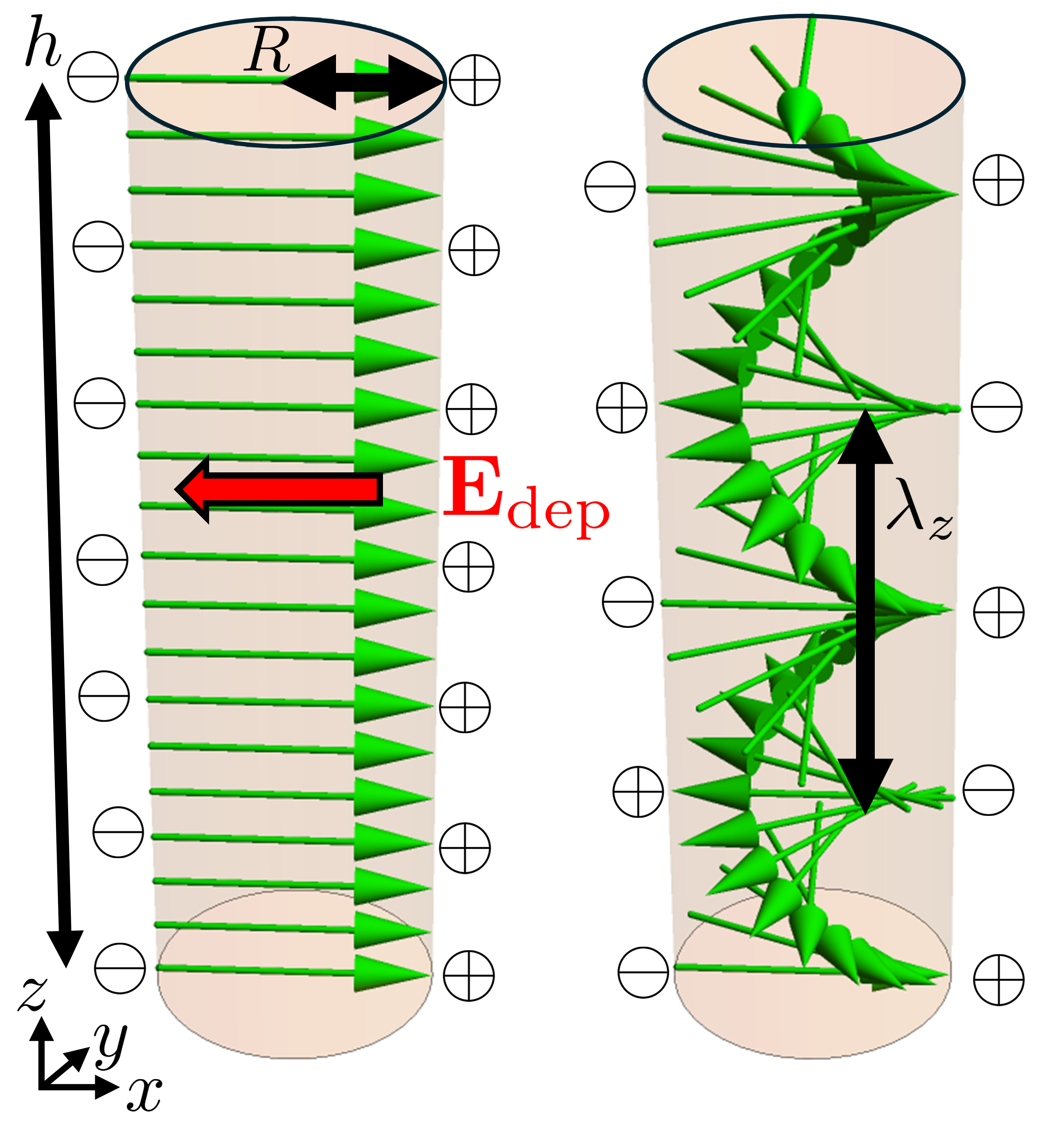}
\caption{\label{fig:twistedcylinder}   \textbf{Cylindrical domain schematic}.  A   uniform polarization $\mathbf{P} = P_0\hat{\mathbf{x}}$ (left panel) incurs an energetic cost due to the uncompensated charges and consequent depolarization field $\mathbf{E}_{\mathrm{dep}}$. Instead, the polarization $\mathbf{P}$ can \textit{twist} (right panel) with a certain period $\lambda_z$ and create regions of alternating charge, thereby partially mitigating the energetic cost of the depolarization field. This twist  is balanced by the elastic cost of the twist. } 
\end{figure}

To begin the theoretical analysis, we first review Khachaturyan's theoretical prediction from 1975 \cite{khachaturyan} that $\mathrm{N}_\mathrm{F}$s   may spontaneously develop twisted polarization $\mathbf{P}$ domains with a certain period $\lambda_z$.
 We   consider a cylindrical domain with radius $R$ and length\ $h$, as shown in Fig.~\ref{fig:twistedcylinder}. In the absence of twist, a uniform polarization ($\mathbf{P} = P_0 \hat{\mathbf{x}}$, say) will generate a strong depolarization field due to the accumulation of uncompensated charge on the domain boundary, as shown on the left panel of Fig.~\ref{fig:twistedcylinder}.
The charges can be partially compensated by twisting the direction of $\mathbf{P}$ along the cylinder, as demonstrated on the right panel of Fig.~\ref{fig:twistedcylinder}.  Note that this model is an idealization, as the polarization $\mathbf{P}$ in a real sample will typically not terminate with any component normal to the boundary, precisely to avoid the accumulation of the bound charge. There will thus likely be a complex deformation of the polarization near such a boundary. Although such configurations are beyond the scope of our analysis, we can use this idealized model to demonstrate the competition between electrostatics and elasticity in a  simple way.  

The $\mathrm{N}_\mathrm{F}$ has a continuously varying polarization vector $\mathbf{P} \equiv \mathbf{P}(\mathbf{r})=P_0 \mathbf{n}(\mathbf{r})$ within the material, with some fixed magnitude $|\mathbf{P}|=P_0$ and a varying orientation $\mathbf{n}(\mathbf{r})$. The bound charge distribution is given by $\rho = - \nabla \cdot \mathbf{P}$. We can calculate the corresponding (screened) electrostatic energy as 
\begin{equation}
F_{\rho}= \frac{1}{8 \pi \epsilon \epsilon_0}\, \iint \, \frac{\rho(\mathbf{r})\rho(\mathbf{r}')e^{-\kappa |\mathbf{r}-\mathbf{r}'|}}{|\mathbf{r}-\mathbf{r}'|} \,\mathrm{d}\mathbf{r}'\,\mathrm{d}\mathbf{r}, \label{eq:electrostatic}
\end{equation}
where $\epsilon$ is the relative dielectric constant of the material and $\kappa \equiv 1/\lambda_D$ is the (inverse) Debye screening length as derived in the Debye-H\"uckel approximation. We have $\kappa \approx  \sqrt{n}\, e/ \sqrt{\epsilon \epsilon_0 k_B T}$ for monovalent ions with concentration $n$, with $e$ the fundamental charge, $\epsilon \epsilon_0$ the material permittivity, and $k_BT$ the thermal energy. This free energy $F_{\rho}$ can be expressed in Fourier space as
\begin{equation}
F_{\rho} =\frac{1}{2 \epsilon \epsilon_0}\,\int\frac{|\mathbf{k} \cdot \tilde{\mathbf{P}}_{\mathbf{k}}|^2}{|\mathbf{k}|^2+\kappa^{2}}   \, \, \,\frac{ \,\mathrm{d}{\mathbf{k}}}{(2\pi)^{3}}, \label{eq:FourierDipole}
\end{equation}
where $\tilde{\mathbf{P}}_{\mathbf{k}}\equiv \int\mathrm{d}\mathbf{r}\,e^{-i\mathbf{k} \cdot \mathbf{r}} \mathbf{P}(\mathbf{r})$. Even in cases where $\rho(\mathbf{r})$ vanishes in the bulk, we may  get contributions to $F_{\rho}$ due to   charges at the boundaries. Note that our expression in Eq.~\eqref{eq:FourierDipole} assumes that the screened Coulomb interaction occurs throughout the system, including outside of the domain with nonzero polarization $\mathbf{P}$.   

Spatial variations of $\mathbf{P}$ will incur elastic energy penalties. 
The (nematic) elastic energy is given by the Frank form
\begin{align}
F & _n= \int \mathrm{d}\mathbf{r} \left[  \frac{K_1}{2} (\nabla \cdot \mathbf{n})^2+\frac{K_2}{2}(\mathbf{n} \cdot (\nabla \times \mathbf{n}))^2   +\frac{K_3}{2}(\mathbf{n} \times (\nabla \times \mathbf{n}))^2 \right], \label{eq:frankfe}
\end{align}
with $K_1$, $K_2$, and $K_3$ the splay, twist, and bend elastic constants, respectively \cite{degennesprost}.  Note that, in general, $\mathbf{P}$ and $\mathbf{n}$ have to be treated separately and are not necessarily colinear. Moreover, the proper order parameter for the nematic portion is a symmetric, traceless tensor $Q$. The more general starting point for describing the $\mathrm{N}_\mathrm{F}$ is given in, e.g., Ref.~\cite{merteljPRE2022}. Here we only look at the competition between elastic distortions of the nematic order and the electrostatic self-interaction, assuming that the nematic order always aligns with $\mathbf{P}$. Any other effects will be incorporated in renormalizations of the constants (e.g., the splay $K_1$ and the dielectric constant $\epsilon$) and we leave a more general analysis for future work. Note also that we are neglecting the saddle-splay elastic contribution which can play a role at sample boundaries or at disclinations \cite{zumersaddlesplay}. In our case, we will have strong anchoring conditions and any singularities will occur at polarization domain boundaries, which we will treat by introducing a phenomenological description of the energetic cost for domain wall formation, which may include effects from the saddle-splay.

We look at $\mathbf{P}$ configurations of the form:\begin{equation}
\mathbf{P} = P_0(\cos[ \phi(z)], \sin [\phi(z)],0)\Theta(x,y), \label{eq:ansatztwist}
\end{equation}
where $0 \leq z \leq h$, $\phi(z)$ is the polar angle of the $\mathbf{P}$ orientation, and  $\Theta(x,y)=1$ whenever $x^2+y^2\leq R$ and $\Theta(x,y)=0$ otherwise. This means that we expect to have uncompensated charges at the cylinder boundary, as shown in Fig.~\ref{fig:twistedcylinder}.  We now assume without loss of generality that the angle $\phi(z)$ is a periodic function with some period $2\pi/k_z$ which might tend toward infinity:
\begin{equation}
\phi(z)=k_z z+\psi(z),
\end{equation} 
where $\psi(z)=\psi(z+2\pi/k_z)$. We may expand the phase factor associated with this angle as
\begin{equation}
e^{i \phi(z)} = e^{i k_z z}\sum_{m=-\infty}^{\infty} A_m e^{i k_z mz}, \label{eq:Fourierseriestwist}
\end{equation}
where $A_m$ are complex Fourier coefficients satisfying\begin{equation}
\sum_{m=-\infty}^{\infty} A_m A^*_{m-n} =\delta_n, \label{eq:orthogonality}
\end{equation}
with $\delta_n=1$ for $n=0$ and $\delta_n=0$ otherwise. Provided we have a thick sample with  $h \gg \lambda_D$, we substitute  Eqs.~(\ref{eq:ansatztwist},\ref{eq:Fourierseriestwist}) into Eq.~\eqref{eq:FourierDipole} and find that\begin{equation}
F_{\rho} = \frac{\pi P_0^2 R^2h}{2 \epsilon \epsilon_0}\,\sum_{n=-\infty}^{\infty} I_1 (\alpha _n)K_1 (\alpha_n)   \left[ A^2_{n}   +(A^*_{n})^2     \right],  \label{eq:dipolartwist}
\end{equation}
where $\alpha_n \equiv \,R \sqrt{ [ k_z(n+1)]^2+ \kappa^2}$ and $I_1(\alpha)$, $K_1(\alpha)$ are the modified Bessel functions of the first and second kind, respectively. It is worth noting that $I_1(\alpha)K_1(\alpha)$ is a monotonically decreasing function of $\alpha$, meaning that the electrostatic energy favors large values of $\alpha_n \sim k_z$.

Given the ansatz in Eq.~\eqref{eq:ansatztwist} and ignoring any elastic deformation or anchoring energy at the cylinder boundary, only the \textit{twist} term proportional to $K_2$ contributes to the elasticity and we find
\begin{equation}
F_n  =\frac{\pi K_2 k_z^2R^2h}{2}+\frac{\pi K_2 R^2h}{2 \lambda_z}\, \, \int_0^{\lambda_z}\mathrm{d}z\,\left(\frac{d\psi}{dz} \right)^2. 
\end{equation}
We may now move to Fourier space by making use of the expansion in Eqs.~(\ref{eq:Fourierseriestwist},\ref{eq:orthogonality}). We find that
\begin{equation}
F_n  = \frac{\pi R^2 hK_2k_z^2}{2}\left[1+    \sum_{n=-\infty  }^{\infty}n^2 |A_{n} |^2\right],
\end{equation}
which  is minimized for  $k_z \rightarrow 0$. Higher order modes $A_n$ with $|n|>0$  cost more elastic energy.  This is less obvious for the electrostatic interaction in Eq.~\eqref{eq:dipolartwist}, but Khachaturyan argues on general grounds \cite{khachaturyan} that there is a stable free energy minimum with  $A_n=0$ for all $|n|>0$.

Looking at solutions with just the $n=0$ mode, we find that the total free energy is given by
\begin{equation}
F=F_n+F_{\rho} = \frac{\pi  R^2h}{2}  \left[\frac{P_0^2I_1 (\alpha_0)K_1 (\alpha_0)}{ \epsilon \epsilon_0} + K_2k_z^2\right], \label{eq:twistresult}
\end{equation}
where $\alpha_0=R \sqrt{  k_z^2+\kappa^2}$ and  $\sum_n |A_n|^2=A_0^2=(A_0^*)^2=1$ from Eq.~\eqref{eq:orthogonality}.
The total free energy in Eq.~\eqref{eq:twistresult} now can be minimized with respect to $k_z$ for $\alpha_0,R/\lambda_D \gg 1$ [so that $I_1(\alpha_0) K_1(\alpha_0) \approx (2\alpha_0)^{-1}$].  We find a minimum free energy at $k_z =k_z^*$, which corresponds to a preferred pitch $\lambda_z\equiv 2\pi/k_z^*$  that reads\begin{equation}
\lambda_z =2 \pi\left[\frac{  P_0^{4/3}}{ (4K_2 R \epsilon \epsilon_0)^{2/3}}- \kappa ^2\right]^{-1/2}. \label{eq:pitch}
\end{equation}
Substituting in reasonable values $\epsilon= 10$, $P_0 =4.4 \times 10^{-2}~\mathrm{C}/\mathrm{m}^2$, $K_2=10~\mathrm{pN}$, and $R = 100~\mu\mathrm{m}$, we find a pitch of $\lambda_z \approx0.4~\mu\mathrm{m}$ assuming no screening $(\kappa=0$) and increasing with increasing $\kappa$. This result is consistent with the previously reported data \cite{sciencetwist2024} and with the observation that twists disappear in our thinnest cells [see Fig.~\ref{fig:Planar}a-c].   Note, however, that $\lambda_z$ in the considered model of an infinitely long cylinder cannot be compared directly to the experimental data obtained in Fig.~\ref{fig:Planar} for flat planar samples of large area since the model does not account for the anchoring effects at the lateral surface of the cylinder. In the experiments, the azimuthal anchoring at the top and bottom plates will typically force the sample to adopt either a uniform structure or a twist by $ \pi$ from the top to bottom of the cell, as discussed below.

 This result for $\lambda_z$ was also recently found by Paik and Selinger using a different analysis \cite{paikselinger}.  Note that, due to screening, the twisted state occurs only when the concentration of ions is below some critical value:
\begin{equation}
c< c^*=\frac{(\epsilon \epsilon_0 )^{1/3}k_B TP_0^{4/3}}{  (4K_2 R  )^{2/3}e^2}. \label{eq:criticalc}
\end{equation}
 Given the reasonable parameters mentioned above and $T=350~\mathrm{K}$, we find  $c^* \approx 5 \times 10^{21}~\mathrm{ions}/\mathrm{m}^{3}$, which is less than the concentration of ions measured in the N phase, $c = (5-6) \times 10^{22}~\mathrm{ions}/\mathrm{m}^{3}$. However, the experimental data obtained in the N phase might not be representative of the concentration of ions in the $\mathrm{N}_\mathrm{F}$ phase.  Also, our analysis so far has not taken into account the cell boundary conditions. Nevertheless, the theoretical estimate appears to be consistent with the experiments in which the TDs disappear when DIO is doped with the ionic fluid BMIM-$\mathrm{PF}_\mathrm{6}$ [see Fig.~\ref{fig:Planar}f], which could   increase the concentration of ions by orders of magnitude.  
 
 \subsection*{Critical cell thickness for $\pi$-twists }
 
 We now consider an $\mathrm{N}_\mathrm{F}$  confined to a  cell with certain imposed nematic orientation $\mathbf{n}_0$ at the top and bottom of the cell, as in the samples shown in  Fig.~\ref{fig:Planar} with strong anchoring. There is a critical thickness $h^*$ for which we get twisted domains if $h>h^*$ and domains with uniform $\mathbf{P}$ orientation for $h<h^*$. When $h>h^*$, the sample thickness $h$ will  set the periodicity of the twist along the $z$ direction (see  Fig.~\ref{fig:Planar}e-j).

 We can calculate the critical thickness $h^*$ by considering a   square domain with dimension $L$. We assume that $\mathbf{P}$ remains in the $xy$ plane and that $\mathbf{P}$ runs along $\hat{\mathbf{y}}$ on the top and bottom surfaces. We compare two polarization configurations: a uniform $\mathbf{P}_{\mathrm{uniform}}=P_0 \hat{\mathbf{y}}$ and a $\pi$-twisted $\mathbf{P}_{\pi-\mathrm{twist}}=P_0 \sin(\pi z/h)\hat{\mathbf{x}}+P_0 \cos(\pi z /h)\hat{\mathbf{y}}$, with all polarizations vanishing outside of the region $-L/2 < x,y<L/2$ and $0<z<h$. The Fourier transforms are
\begin{equation}
\begin{cases}\tilde{\mathbf{P}}_{\mathrm{uniform}} =\dfrac{8 P_0  \sin\left( \frac{k_z h}{2}\right)\sin\left( \frac{k_x L}{2}\right)\sin\left( \frac{k_y L}{2}\right)\hat{\mathbf{y}}}{ k_x k_yk_z  e^{\frac{ih k_z}{2} }}  \\[8pt]
\tilde{\mathbf{P}}_{\pi-\mathrm{twist}}=\dfrac{8P_0  h\sin\left( \frac{k_x L}{2}\right)\sin\left( \frac{k_y L}{2}\right)\cos\left( \frac{h k_z}{2} \right)[\pi \hat{\mathbf{x}}+ihk_z\hat{\mathbf{y}}]}{  k_x k_y[\pi^2-h^2k_z^2 ]e^{\frac{ih k_z}{2} }}
\end{cases} . \label{eq:pitwistfourier}
\end{equation}
 Substituting Eq.~\eqref{eq:pitwistfourier} into Eq.~\eqref{eq:FourierDipole} and assuming that $L \kappa \gg 1$ (strong screening or large sample size)
yields the dipolar energy
\begin{equation}
F_{\rho}= \begin{cases} 
\dfrac{L hP_0^2}{\epsilon \epsilon_0 \kappa}& \mbox{for }\tilde{\mathbf{P}}_{\mathrm{uniform}}\\[10pt]
\dfrac{   LhP_0^2}{  4\epsilon \epsilon_0 \kappa   }& \mbox{ for }\tilde{\mathbf{P}}_{\pi-\mathrm{twist}}
\end{cases}. \label{eq:pitwistE}
\end{equation}
The electrostatic energy cost of the polarization configuration gets a four-fold decrease from the $\pi$-twist along the $z$-axis. 

The $\pi$-twisted configuration incurs an elastic energy penalty given by $F_n=K_2 L^2 \pi ^2/(2h)$, which follows from substituting $\mathbf{n}_{\pi-\mathrm{twist}}=  \sin(\pi z/h)\hat{\mathbf{x}}+  \cos(\pi z /h)\hat{\mathbf{y}}$ into the Frank free energy, Eq.~\eqref{eq:frankfe}.  The balance between elastic and electrostatic energies yields a critical thickness
\begin{equation}
h^*=\frac{\pi}{P_0} \sqrt{\frac{2\epsilon \epsilon_0 \kappa K_2 L }{3  }    }.
\end{equation}
Even for   large domains with $L \approx 1~\mathrm{cm}$, we find a   small $h^* \approx 0.2-2~\mu\mathrm{m}$, for a wide range of screening lengths $\lambda_D =\kappa^{-1} \approx0.01-1~\mu\mathrm{m} $.   In the experiments, we find that essentially all domains are twisted for thicknesses $h>2~\mu\mathrm{m}$, consistent with this result. Confinement can induce chirality (twist) in solid state ferroelectrics, as well, especially in  nanostructured materials \cite{Lukyanchuk1, Lukyanchuk2}, although intrinsically chiral solid ferroelectrics are also possible.

\subsection*{Model of stripe domain patterns}

 \begin{figure}[htp]
\centering
\includegraphics[width=0.4\textwidth]{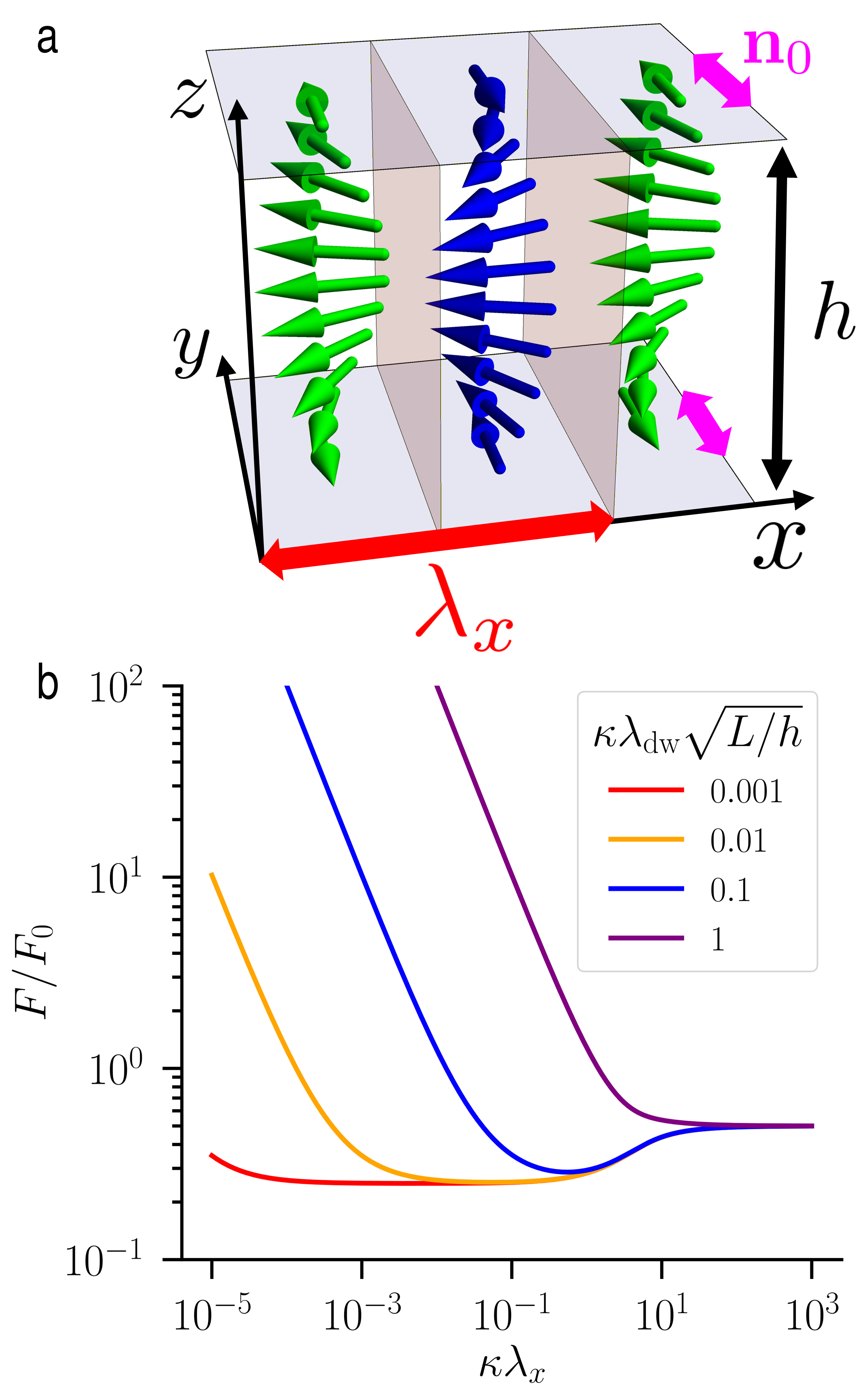}
\caption{\label{fig:square2} \textbf{Striped domain model.}     \textbf{a} Schematic of a possible polarization direction in the striped domains with characteristic wavelength $\lambda_x$, with the anchoring nematic direction $\mathbf{n}_0$ indicated on the top and bottom surfaces. The gray surfaces separating the blue and green domains could be either disclination lines, solitons, of fixed-width domain walls spanning the entire cell height $h$.   \textbf{b} Dimensionless free energy [Eq.~\eqref{eq:stripeenergy}] of the striped domain configuration, taking into account both the electrostatic energy contribution and the energy of domain walls (captured by the parameter $\lambda_{\mathrm{dw}}$). We see that there is a minimum at a certain $\kappa \lambda_x$, indicating that the striped configuration is favorable. We show the energy for various values of the dimensionless parameter $\kappa \lambda_{\mathrm{dw}} \sqrt{L/h}$ discussed in the text.} 
\end{figure}

The domains   in Fig.~\ref{fig:Planar} are reminiscent of structures found in thin films of solid, uniaxial ferroic materials \cite{thinfilmreview}, where the polarization $\mathbf{P}$  forms stripes of alternating orientation (although the polarization typically has components perpendicular to the film surface, unlike the $\mathrm{N}_\mathrm{F}$ which has $\mathbf{P}$ parallel to the $xy$ plane) \cite{hardstripe1,hardstripe2}.
The stripes in solid ferroics are generated to mitigate the  depolarization field \cite{kittelstripe,elderstripe}. Similar striped patterns appear in ferromagnetic crystals, with the patterns analyzed 90 years ago by Landau and Lifshitz \cite{LLstripes}.   

Consider a thin, square cell with thickness $h$ and square cross section with area $A_{xy}=L^2$, where $L=L_x=L_y$ is the linear extent of our sample. The basic idea, also discussed in detail by Kittel for solid ferromagnetic materials \cite{kittelstripereview},  is that the dipolar energy density $f_{\rho}=F_{\rho}/A_{xy}$ per area of the cell will scale approximately as $f_{\rho} \propto \lambda_x$, with $\lambda_x$ the stripe size for stripes running along the $y$-axis, say. Meanwhile, the introduction of alternating domains of $\mathbf{P}$ will incur a cost due to the domain walls. There will be approximately $2L/\lambda_x$ domain walls in the sample, so the associated free energy density will scale according to $f_{\mathrm{dw}} \propto 1/ \lambda_x$. There should be an optimal value of $\lambda_x$ which balances the two energy densities $f_{\mathrm{dw}}$ and $f_{\rho}$.

Let us analyze the optimal wavelength $\lambda_x$ by assuming that we have very strong anchoring so that each stripe satisfies the bidirectional boundary conditions at the cell surface, as illustrated in Fig.~\ref{fig:square2}a.   One possibility, illustrated in Fig.~\ref{fig:square2}a, is that adjacent domains are of opposite chirality. The corresponding polarization field $\mathbf{P}_{\mathrm{stripe}}$ for such a configuration reads
\begin{equation}
\mathbf{P}_{\mathrm{stripe}}=\frac{4 P_0}{\pi } \, \sum_{n=1}^{\infty}  \, \frac{1}{n} \, \sin\left( \frac{\pi n}{2}\right)\cos (n q_xx)\cos\left( \frac{\pi z}{h}\right) \hat{\mathbf{y}}  +P_0\sin\left( \frac{\pi z}{h}\right)\hat{\mathbf{x}}, \label{eq:stripeP}
\end{equation}
where $q_x=2\pi/\lambda_x$ is the wavevector associated with the stripe wavelength $\lambda_x$ (see red double arrow in Fig.~\ref{fig:square2}a). The summation over $n$ is the mode expansion of a square wave, so that we have a rapid reorientation of $\mathbf{P}$ from $+P_0 \hat{\mathbf{y}}$ to $-P_0\hat{\mathbf{y}}$ at the top and bottom of the cell ($z=0,h$), as illustrated in Fig.~\ref{fig:square2}a.

A sudden jump in the polarization ($\pm P_0 \hat{\mathbf{y}}$ in Fig.~\ref{fig:square2}a) is   unrealistic and the detailed structure near the flip (gray surfaces separating the green and blue arrows in Fig.~\ref{fig:square2}a)   could take a variety of forms including  disclination line pairs, a solitonic structure, or a fixed-width domain wall, as discussed in more detail below.   We assume here that this region does not significantly influence the electrostatic contribution to the energy, instead contributing   to the elastic cost of the polarization domain. We also assume that the striped configuration  has some lateral extent $0<x,y<L$, where $L$ is some integer multiple of $\lambda_x$, for simplicity.    Although the system is overall charge neutral, there will be   regions of charge at the boundary of the domains along the $x,y$ directions where the polarization arrows terminate, creating depolarization fields. 

We substitute the Fourier-transformed Eq.~\eqref{eq:stripeP} into Eq.~\eqref{eq:FourierDipole} to find the dipolar energy of this configuration, again assuming a screened Coulomb interaction everywhere with a screening length $\kappa^{-1}$.   We find, for large sample sizes $L \kappa \gg 1$, that
\begin{align}
F_{\rho}&= \frac{4P_0^2  }{\pi^3  \epsilon \epsilon_0}\,\int\frac{\cos^2\left(\frac{z}{2}\right) \sin^2\left (\frac{x L}{2 h} \right)\sin^2\left( \frac{  yL}{2 h} \right)}{( x^2+y^2+z^2+h^2\kappa^2)(\pi^2 -z^2)^2} \left[   \frac{16  x^2z^2}{\pi^2h } \left[ \sum_{n=1}^{\infty}    \frac{ h^2\lambda_x^2    \sin\left( \frac{\pi n}{2}\right) }{n( \lambda_x^2x^2-4\pi^2n^2h^2)} \right]^2+ \frac{ h^3\pi^2     }{ y^2}\right]   \mathrm{d}x\,\mathrm{d}y\,\mathrm{d}z \nonumber \\
  & \approx\frac{h^2 LP_0^2  }{\pi^3 \epsilon \epsilon_0  }\,   \left[  \int_{-\infty}^{\infty}  \mathrm{d}z\,   \frac{\cos^2\left(\frac{z}{2}\right)  }{[\pi^2-z^2]^2}   \frac{\pi^4}{   \sqrt{ z^2+h^2\kappa^2}}+\frac{2 \pi \lambda_x}{h}  \sum_{n=1}^{\infty}\, \frac{\sin^2\left(\frac{\pi n}{2} \right)}{n^2 \sqrt{(2 \pi n)^2+ (\kappa \lambda_x)^2}}    \right].  \label{eq:electroEstripe}
\end{align}
 As long as the screening contribution in the sum in Eq.~\eqref{eq:electroEstripe} is negligible ($\lambda_x \kappa \lesssim 1$), then the electrostatic  energy $F_{\rho}$ grows with $\lambda_x$.

The nature of the domain walls may be complex due to the twist in the polarization $\mathbf{P}$. Experiments indicate that the domain wall may consist of surface disclination lines, with complex elastic distortions in the bulk near the disclinations \cite{sciadv2021domainwall,QWeiPNAS2024}. In addition, these walls will tend to have vanishing divergence of the polarization $\mathbf{P}$, so we expect no bound charges at the wall locations \cite{chen2020first,FNconics2024}. An estimate for the energy of the wall (per unit length) due solely to  disclinations would be $f^{\mathrm{disc}}_{\pi} \approx 2 K$, with $K$ an elastic constant. Another possibility is that the domain wall is a solitonic, Bloch wall structure with an energy governed by the anchoring conditions, which we may call a $\pi$-wall/soliton as $\mathbf{P}$ flips by 180${}^{\circ}$ across the wall \cite{solitons2022}. The energy per unit length of one such $\pi$-soliton is given by $f^{\mathrm{soliton}}_{\pi}\approx 2\sqrt{2 K h W}$, where $K$ is an elastic constant, $W$ is the anchoring strength, and $h$ is the cell thickness.   A third possibility is that the energy of the wall is proportional to the cell thickness, which may occur if the elastic distortion of the wall extends across the entire thickness $h$, which may occur for thin cells, so that $f_{\pi}^{\mathrm{wall}} \approx K h/\lambda_{\mathrm{wall}} $, where $\lambda_{\mathrm{wall}}$ is a characteristic width of the wall. We assume here that the characteristic width is independent of $h$. We will call this case the  fixed-width domain wall. In summary, the total elastic cost of domain walls is   approximately $F_{\mathrm{dw}} \approx 2 f_{\pi} L^2/\lambda_x$, with $f_{\pi}$ either the disclination line pair $(f_{\pi}=f_{\pi}^{\mathrm{disc}})$, $\pi$-soliton ($f_{\pi}=f^{\mathrm{soliton}}_{\pi}$), or the fixed-width wall ($f_{\pi}=f_{\pi}^{\mathrm{wall}} $)  energy density. More complicated possibilities or combinations of these three cases are also possible, but we will focus on just these three cases for simplicity.

Putting it all together, the total energy $F=F_{\rho}+F_{\mathrm{dw}}$ of a striped domain configuration reads
\begin{align}
\frac{F}{F_0}& =\frac{1     }{4  }+ \frac{\kappa \lambda_x}{\pi^2 } \sum_{n=1}^{\infty}\, \frac{[1-(-1)^n]}{n^2  \sqrt{(2\pi n)^2 +(\kappa \lambda_x)^2}}  + \, \frac{(\kappa \lambda_{\mathrm{dw}})^2L}{    \kappa \lambda_xh}  , \label{eq:stripeenergy}
\end{align}
where $F_0= hP_0^2 L(\epsilon \epsilon_0 \kappa)^{-1}$ is a characteristic free energy and $\lambda_{\mathrm{dw}}=(2f_{\pi} \epsilon \epsilon_0)^{1/2}P_0^{-1}$ is a characteristic length associated with the domain wall energy density $f_{\pi}$.  A similar length was derived some time ago by balancing analogous energetic contributions \cite{clarknote}.
 We have made an additional assumption that we can take $h \kappa \gg 1$ in the left-over integration in Eq.~\eqref{eq:electroEstripe}, which will not change the location of the minimum of $F$ with respect to $\lambda_x$. 
 We plot  the dimensionless free energy $F/F_0$ in  Eq.~\eqref{eq:stripeenergy} (using the first 1000 non-zero terms in the sum) versus $\kappa \lambda_x$ in Fig.~\ref{fig:square2}b for various values of $\kappa \lambda_{\mathrm{dw}} \sqrt{L/h}$, with $L$ the linear extent of the sample. Numerically minimizing the function in Eq.~\eqref{eq:stripeenergy}  demonstrates that for $\kappa \lambda_{\mathrm{dw}} \sqrt{L/h}< 0.62$, the free energy curve has a global minimum at a finite value of $ \lambda_x$.
For larger values, the free energy minimum corresponds to $\lambda_x \rightarrow \infty$ and a uniform polarization state. In the low screening limit $\kappa \lambda_{\mathrm{dw}} \ll \sqrt{h/L}$, the preferred wavelength is at  
\begin{equation}
\lambda_x^* \approx 5.4\lambda_{\mathrm{dw}} \sqrt{\frac{L}{h}}. \label{eq:lowkappa}
\end{equation}

\subsection*{Model of pie-slice domains}

   \begin{figure}[htp]
\centering
\includegraphics[width=0.4\textwidth]{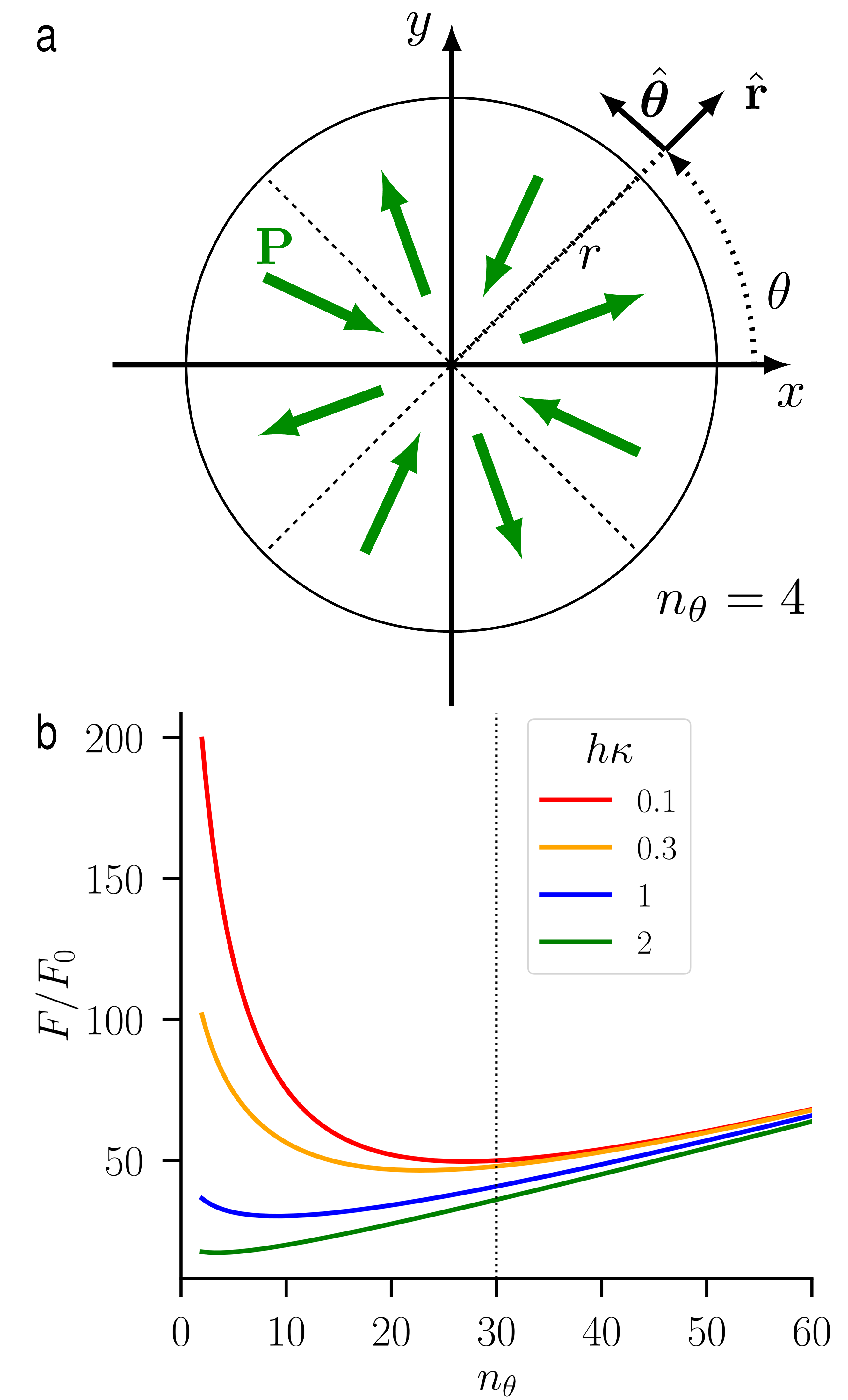}
\caption{\label{fig:radials} \textbf{Pie-slice domain model.}  \textbf{a} Schematic of the polarization orientation near the cell surface for a cell pre-patterned with a $+1$ defect. Apart from the domain walls (along the $x$ and $y$ axes and along the dashed line), the polarization $\mathbf{P}$ runs along the radial direction $\hat{\mathbf{r}}$. The polarization configuration has $n_{\theta}=4$ pairs of pie-slice sectors of  anti-parallel $\mathbf{P}$. \textbf{b} Plot of the rescaled free energy $F/F_0$ in Eq.~\eqref{eq:fullenergy} of a pie-slice pattern with $n_{\theta}$ cuts for various values of $h\kappa$, with $\kappa$ the inverse Debye screening length and $h$ the cell thickness. We fix $ R/h =100$ and $\eta_{\theta}=30$. We see that the Coulomb and elastic energies balance to generate an optimal value of $n_{\theta}$ (the free energy minimum). The dashed vertical line indicates   the optimal value, $\eta_{\theta}$, in the low screening $\kappa \rightarrow 0$ limit.    } 
\end{figure}

We now consider the $+1$ aster defect anchoring condition, with $\mathbf{n}_0 = \hat{\mathbf{r}}$ the preferred orientation at the top and bottom cell surfaces. A pure $+1$\ aster defect in the polarization vector, $\mathbf{P}=P_0 \hat{\mathbf{r}}$, has a corresponding bound charge density distribution
\begin{equation}
\rho(r)=-\nabla \cdot \mathbf{P}=-\frac{P_0 }{r}.
\end{equation}
The distribution of bound charge at the defect is energetically costly, so the $\mathrm{N}_\mathrm{F}$ prefers to reorient along the $\hat{\bm{\theta}}$ direction to create a net neutral charge configuration, forming a series of $n_{\theta}$ pairs  of alternating $\mathbf{P}$ domains (with $\mathbf{P} \parallel \pm \hat{\mathbf{r}}$), as shown in Fig.~\ref{fig:radials}(a).  We also expect a $\pi$-twist along the $z$-direction, but we will neglect the $z$-dependence for simplicity, focusing on the SDs (see Fig.~\ref{fig:SD}).

Consider the anchoring-imposed splay polarization $\mathbf{P}$ configuration in a circular domain of radius $R$. Far from the defect, as $R \rightarrow \infty$, the polarization will be uniform and we  expect to see striped domains with the preferred wavelength $\lambda_x^*$ calculated in the previous section. Near the defect, we expect the bound charge distribution in the bulk should play a more significant role.  Assuming the polarization remains in the $xy$ plane and does not depend on the distance $r$ from the defect, the   form of the polarization vector of such a configuration is
\begin{equation}
\mathbf{P}= P_0 \cos[\phi(\theta)]\hat{\mathbf{r}}+P_0\sin[\phi(\theta)]\hat{\bm{\theta}}  ,
\end{equation}
where $\phi(\theta)$ describes the polarization orientation away from the $\hat{\mathbf{r}}$ direction. The bound charge distribution due to this polarization $\mathbf{P}$   is
\begin{equation}
\rho(r,\theta)\approx -\frac{P_0  \cos[\phi(\theta)]}{r} , \label{eq:rhotheta}
\end{equation}
where we have assumed that $\partial_{\theta} \phi \ll 1$ is negligibly small throughout most of the sample. Regions of opposite (cancelling) charge are created by alternating between $\phi(\theta)=0,\pi$. Note that this approximation will break down close to the $+1$ defect (the small $R$ region)  where all of the domain tips collide.  The defect structure there is complex and, even in the absence of domains, the $+1$ defect splits into two $+1/2$ defects joined by a domain wall, as discussed previously (see Fig.~\ref{fig:SD}h). We   thus consider intermediate values of $R \approx 100~\mu\mathrm{m}$ when comparing to experiments in the next section.

 We  perform a single-mode approximation so that $\cos[\phi(\theta)] \approx \cos(n_{\theta} \theta)$.   Then, in cylindrical coordinates, the screened Coulomb potential  is given by
\begin{equation}
\frac{ e^{-\kappa |\mathbf{r}-\mathbf{r}'|}}{|\mathbf{r}-\mathbf{r}'|} =\frac{2}{\pi}\sum_{n=0}^{\infty} (2-\delta_{n})\cos [n(\theta-\theta')] \  \int_0^{\infty} \mathrm{d}k\,I_n(x_<)K_n(x_>)\cos[k(z-z')], \label{eq:coulombcylinder}
\end{equation}
where  $\delta_n=1$ if $n=0$ and $\delta_n=0$, otherwise. Also, $x_{<,>} = \sqrt{k^2+\kappa^2}\,r_{<,>}$ and $r_<$ ($r_>$) is the smaller (larger) of the polar distances $r$ and $r'$. Substituting $\rho(\theta)$ from Eq.~\eqref{eq:rhotheta} and Eq.~\eqref{eq:coulombcylinder} into Eq.~\eqref{eq:electrostatic} yields (after some algebra and an identity for the integral of a single modified Bessel function of the first kind $I_{\nu}(z)$ \cite{gradshteyn}) \
\begin{equation}
F_{\rho}    = \frac{ 32\pi^2h^3P_0^2 }{  \epsilon \epsilon_0} \sum_{m=0}^{\infty} (-1)^m\int_0^{\infty}\mathrm{d}u\,\frac{  \sin^2 (\frac{u}{2})}{u^2(u^2+h^2\kappa^2)}   \, \int_0^{ \frac{R}{h}\sqrt{u^2+h^2\kappa^2}} \mathrm{d}v  \,\,I_{2m+1+n_{\theta}}(v)K_{n_{\theta}}(v), \label{eq:exactcoulombpieslice}
\end{equation}
where $K_{\nu}(x)$ is a modified Bessel function of the second kind. The dominant term in the summation is $m=0$ and we can make use of the approximations $I_{n_{\theta}+1}(v)K_{n_{\theta}}(v) \approx v[4n_{\theta}(n_{\theta}+1)]^{-1} ,(2v)^{-1}$ for $v \ll n_{\theta}$ and $v \gg n_{\theta}$, respectively.

The total length of domain wall in the system is $2n_{\theta} R$ so that the total elastic cost is  $F_{\mathrm{dw}}=2n_{\theta}Rf_{\pi}$, with $f_{\pi}$ the domain wall linear energy density  (e.g., either the soliton, the domain wall, or the disclination line pair). We find that the total free energy $F=F_{\rho}+F_{\mathrm{dw}}$  of the   pie-slice configuration with $n_{\theta}$ cuts (i.e., $2n_{\theta}$ pie slices)  reads
\begin{equation}
\frac{F}{F_0}=\frac{8h\eta_{\theta}^2 }{R}  \ \int_0^{\infty}\mathrm{d}u\,  \frac{ \sin^2 (\frac{u }{2 })\, \Xi_{n_{\theta}}(\frac{R}{h}\sqrt{u^2 +\kappa^2h^2 })}{  u^2(u^2+\kappa^2h^2 )}+ n_{\theta}, \label{eq:fullenergy}
\end{equation}
where $F_0 =2 Rf_{\pi} $ is a characteristic energy,   $\eta_{\theta}= 2\pi h  / \lambda_{\mathrm{dw}}$ is the  optimal number of sectors without screening  $\eta_{\theta}=n_{\theta}^* (\kappa \rightarrow 0)$, and $\Xi_{n_{\theta}}(z) \equiv \int_0^zI_{n_{\theta}+1}(z)K_{n_{\theta}}(z)\,\mathrm{d}z$. The parameter $\lambda_{\mathrm{dw}}\equiv \sqrt{2 f_{\pi} \epsilon \epsilon_0}/P_0$ depends on    $f_{\pi}$ just as in the stripe case. The plot of   Eq.~\eqref{eq:fullenergy} (with numerical evaluation of the integrals) is shown in Fig.~\ref{fig:radials}b. The total free energy exhibits a minimum at a non-zero value of $n_{\theta}$.

We consider two cases: $\kappa R \ll n_{\theta}$ (weak screening) and $\kappa R \gg n_{\theta}$ (strong screening). We find [using asymptotic expansions of the integrals in Eq.~\eqref{eq:exactcoulombpieslice} \cite{sidihoggan}] that, for thin cells compared to the domain size, $h \ll R$,  
\begin{equation}
F_{\rho} \approx\frac{\pi^2 hP_0^2}{\epsilon \epsilon_0}    \begin{cases}
\dfrac{  hR  }{  2 n_{\theta}}& \kappa R \ll n_{\theta} \\[12pt]
\dfrac{ 4\pi }{   \kappa ^2}  \ln \left(\dfrac{  \kappa  R }{  n_{\theta}} \right)& \kappa R \gg n_{\theta}
\end{cases} \label{eq:Coulombpie}
\end{equation}
In both cases, this Coulomb energy  decreases with  increasing $n_{\theta}$, in contrast to the elastic and anchoring energies which will increase proportionally to $n_{\theta}$.
 We   minimize the total free energy $F$ to find the approximate result:
\begin{equation}
 n_{\theta}^* \approx  \begin{cases}
\dfrac{2\pi  h }{\lambda_{\mathrm{dw}} }& \kappa R \ll n_{\theta}^*  \\[12pt]
\dfrac{  4\pi^3 h  }{        R (\kappa  \lambda_{\mathrm{dw}})^2  } & \kappa R \gg n_{\theta}^* 
\end{cases},  \label{eq:pieslicebaretheory}
\end{equation}
  where the large versus small screening conditions have to be checked self-consistently. The value $2n_{\theta}^*$  gives the ``optimal'' number of sectors (i.e., $\pm \hat{\mathbf{r}}$ polarization domains).

\subsection*{Quantitative comparisons between models and experiments}

 \begin{figure}[htp]
\centering
\includegraphics[width=0.4\textwidth]{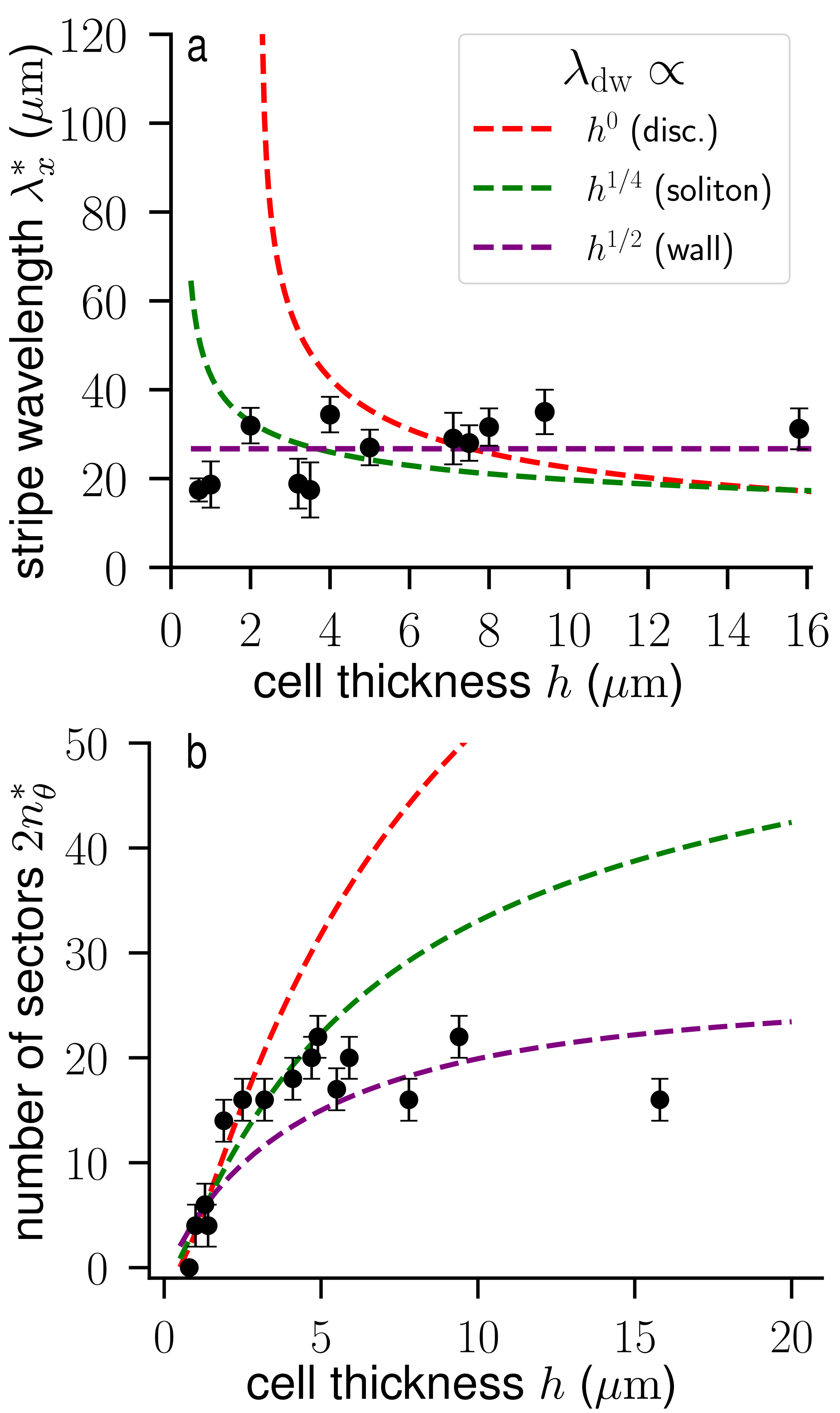}
\caption{\label{fig:experiment}   \textbf{Quantitative comparison.} Experimental (points) and theoretical predictions (dashed lines) for the number of domains as a function of cell thickness $h$. In the theoretical models, we fix    $\lambda_D=\kappa^{-1}=10~\mu\mathrm{m}$ throughout.  \textbf{a} TD stripe wavelength $\lambda_x^* \approx 2 \langle w_{\mathrm{stripe}} \rangle$  calculated form the average stripe width across a sample with unidirectional, bipolar anchoring. The error bars show the standard deviation over observed stripes. The dashed curves are the  wavelengths found by minimizing Eq.~\eqref{eq:stripeenergy} over $\kappa \lambda_x$ for various models of the domain wall cost, which impacts the characteristic length $ \lambda_{\mathrm{dw}}= \sqrt{2\epsilon \epsilon_0 f_{\pi}}/P_0$.   We set $\lambda_{\mathrm{dw}}=0.12~\mu\mathrm{m}$ constant for the disclination line case (red curve). We set $\lambda_{\mathrm{dw}}=(0.0608~\mu\mathrm{m}^{3/4}) h^{1/4}$ for the soliton (green), and  $\lambda_{\mathrm{dw}}=(0.0437~\mu\mathrm{m}^{1/2})h^{1/2}$ for the fixed-width wall (purple).  \textbf{b} The number of pie-slice sectors (SDs) $2n_{\theta}^*$ counted out at a distance  of $R=104~\mu\mathrm{m}$ from the defect center of a cell pre-patterned with a $+1$ ``aster'' defect. The theory curves are found by minimizing the free energy in Eq.~\eqref{eq:fullenergy} with respect to $n_{\theta}$. We again consider the disclination model with $\lambda_{\mathrm{dw}}=1.5~\mu\mathrm{m}$ constant, a soliton with $\lambda_{\mathrm{dw}}=(1.4~\mu\mathrm{m}^{3/4}) h^{1/4}$, and the fixed-width wall with $\lambda_{\mathrm{dw}}=(1.3~\mu\mathrm{m}^{1/2})h^{1/2}$. The error bars are estimated from the uncertainty in the domain count in the sample. } 
\end{figure}

 A key question is how the domain wall energy density $f_{\pi}$ depends on the cell thickness $h$. One possibility is that the walls are disclination  pairs, so that $f_{\pi} \sim 2K \sim 10-1000~\mathrm{pN}$ is independent of the thickness $h$ and is proportional to some combination of elastic constants, here indicated as $K$. However, for smaller thicknesses $h$, it is likely that the elastic distortion at the wall extends through the entire cell thickness $h$ and contributes meaningfully to the energy of the domain wall. In this case, we can consider other scalings such as $f_{\pi} \propto \sqrt{h}, h$ for the soliton and fixed-width domain wall, respectively. These various scalings, then, will change how the characteristic length  $\lambda_{\mathrm{dw}}=(2f_{\pi} \epsilon \epsilon_0)^{1/2}P_0^{-1}$ varies with $h$, with $\lambda_{\mathrm{dw}} \propto h^0, h^{1/4}, h^{1/2}$ for the disclinations, soliton, and fixed-width domain wall, respectively. These three possibilities are shown in Fig.~\ref{fig:experiment} for both the stripe patterns and the pie-slice sector results, which we  consider in more detail below.

The screening length $\lambda_D=\kappa^{-1}$    is not necessarily constant throughout the sample. For example, in the stripe case (UDs and TDs), the screening effects would be most important near the domain or cell boundary. On the other hand, for the radial slices (SDs), the screening near the $+1$ defect would be more relevant.  The value of $\kappa$ may also change with the cell thickness $h$, as ions may penetrate the sample less when the thickness $h$ is less than the screening length $\kappa^{-1}$. We will fix $\kappa=0.1~\mu\mathrm{m}^{-1}$  in our analysis, which seems to work well for the samples considered here. This is a relatively large screening length ($\lambda_D=\kappa^{-1}=10~\mu\mathrm{m}$), but not necessarily large enough that we can take the zero screening limit. 

 For the stripe patterns (TDs), we measure the stripe widths $w_{\mathrm{stripe}}$ across a cell over a distance of about $500~\mu\mathrm{m}$ (see Fig.~\ref{fig:Planar}a-d). The average $\langle w_{\mathrm{stripe}} \rangle $ and standard deviation $ \sqrt{\langle w_{\mathrm{stripe}} ^2\rangle-\langle w_{\mathrm{stripe}}\rangle^2} $  for various cell thicknesses  are shown as the data points and error bars, respectively, in Fig.~\ref{fig:experiment}a. To compare to theory, we minimize Eq.~\eqref{eq:stripeenergy} with respect to $ \lambda_x$  to find the optimal value $\lambda_x^*$ for various values of $h$ and fixed $\lambda_D=\kappa^{-1}=10~\mu\mathrm{m}$ and full sample size $L =1~\mathrm{cm}$. We also vary the value of $\lambda_{\mathrm{dw}}$ to find a favorable match to the experimental data, trying different possibilities for the $h$-dependence: The  red dashed line in Fig.~\ref{fig:experiment}a corresponds to a constant $   \lambda_{\mathrm{dw}} =0.12~\mu\mathrm{m}$, which would be consistent with a domain wall consisting of disclination lines. The green dashed line in Fig.~\ref{fig:experiment}a  corresponds to a solitonic domain wall with $f_{\pi}=f_{\pi}^{\mathrm{soliton}} \propto \sqrt{h}$, for which we set $\lambda_{\mathrm{dw}}=(0.0608~\mu\mathrm{m}^{3/4})h^{1/4}$, a scaling which compares favorably with the data for intermediate thicknesses $h$ between 2 and 8 microns. Finally, if the domain wall elastic distortion extends across the entire cell thickness and we have $f_{\pi}=f_{\pi}^{\mathrm{wall}} \propto h$, then we find that we get good agreement with the data for  $\lambda_{\mathrm{dw}}=(0.0437~\mu\mathrm{m}^{1/2})h^{1/2}$, as shown by the purple dashed line in Fig.~\ref{fig:experiment}a.  Note that in this case, the stripe wavelength $\lambda_x^*$ does not change with the thickness $h$, which is consistent with the experimental results. We may thus conclude that, in our samples, the domain walls between the polarization domains likely have an energetic cost that scales proportionally to the cell thickness $h$.

For the pie-slice domains in the $+1$ aster defect cell, we count the number of domains out at a radius $R = 104~\mu\mathrm{m}$ away from the center of the defect. We then compare to the theoretical result for $2n_{\theta}^*$ given by the minimum of the energy in  Eq.~\eqref{eq:fullenergy} with respect to $n_{\theta}$.  The red dashed curve in Fig.~\ref{fig:experiment}b has a  constant $   \lambda_{\mathrm{dw}} =1.5~\mu\mathrm{m}$,  corresponding to  disclinations as the domain walls. This clearly underestimates the energetic cost of the domain walls, especially for $h>4~\mu\mathrm{m}$. On the other hand, if we consider the solitonic wall (green curve in Fig.~\ref{fig:experiment}b) and set $\lambda_{\mathrm{dw}}=(1.4~\mu\mathrm{m}^{3/4}) h^{1/4}$, we get a better match to the data over a larger range of  $h$. Finally, for the fixed-width domain walls with energies proportional to $h$, we find a good match to the data with $\lambda_{\mathrm{dw}}=(1.3~\mu\mathrm{m}^{1/2})h^{1/2}$, shown by the purple dashed curve in Fig.~\ref{fig:experiment}b. Thus, like in the striped pattern case, we conclude here that the cost of the domain walls between polarization domains likely scales linearly with the cell thickness $h$. However, we note that we have fixed $\kappa=0.1~\mu\mathrm{m}^{-1}$ to a constant and $\kappa$ may well be $h$-dependent. Moreover, the $h$-dependence of $\lambda_{\mathrm{dw}}$ might be more complex than the cases we have considered here. Nevertheless, we note the good agreement between the theoretical results with fixed-width domain wall energy $f_{\pi}=f_{\pi}^{\mathrm{wall}} \propto h$ and the experimental data for both the stripe wavelength calculation and the calculated number of pie slices in Fig.~\ref{fig:experiment}a,b, respectively.

 In summary, the theoretical results for the stripe domain (TD) and  pie slice (SD) patterns match favorably with experimental data, as shown in Fig.~\ref{fig:experiment}a,b, respectively. The two cases seem to have different values for the characteristic domain wall length $\lambda_{\mathrm{dw}}$, with the SD\ domains consistent with a larger (approximately 10 to 30-fold) value of $\lambda_{\mathrm{dw}}$.  This could be due to different values of the effective dielectric constant $\epsilon$, or it could be that the two cases have different values of $\kappa$, which we kept constant here at  $\kappa=0.1~\mu\mathrm{m}^{-1}$ for both cases. Changing the value of $\kappa$ would result in different values of $\lambda_{\mathrm{dw}}$ when matching to experimental data. Variations in $\kappa$ and $\lambda_{\mathrm{dw}}$ could be due to the different nature of the patterns: the stripes are generated due to uncompensated bound charges at the sample boundaries while the radial pie slices form due to a non-vanishing bound charge density at the $+1$ defect center. Free ions may be able to screen the charge better at the defect center, leading to larger $\kappa$ values and, similarly, larger values of $\lambda_{\mathrm{dw}}$ (which is proportional to $\sqrt{\epsilon}$) for the SD case relative to the TD pattern.

\section*{Discussion}

    The long-range Coulomb interactions  for regions with uniform $\mathbf{P}$ generate a large electrostatic energy penalty at region boundaries.  Overall, the Coulomb interactions tend to create regions of opposing  $\mathbf{P}$ directions. On the other hand,   reorientations of the polarization direction  incur elastic energy costs. We have demonstrated that the competition between these two effects creates a variety of domain patterns that depend strongly on the cell thickness and the ionic content.

In micron-thin $\mathrm{N}_{\mathrm{F}}$ cells with the same uniform direction of apolar anchoring at the top and bottom plates, the patterns are formed by striped domains with a uniform polarization that flips by $\pi$ in transition from one domain to the next, see Figs.~\ref{fig:Planar} a-d, \ref{fig:SD}a,b, \ref{fig:BMIM}a, and \ref{fig:radials}a. In thicker cells, the stripes show left- and right-handed $\pi$ twists of polarization around the cell normal, see Figs. \ref{fig:Planar}e-j, \ref{fig:TD}, and \ref{fig:square2}a.     We also considered   cells pre-patterned with a $+1$ radial aster defect. Here,   the system breaks up into   pie-slice  domains due to the bound charge distribution $\rho = - \nabla \cdot \mathbf{P} \propto 1/r$, decaying with distance $r$ from the defect core. The theoretically predicted number of pie slices and the stripe width are consistent with the experiment and with the idea that the screening effect in the pre-patterned splay is stronger than in the case of uniform surface anchoring.

In the future, we hope to test the theory more stringently by  systematically varying the free ion concentration (and $\kappa$, consequently). It would also be interesting to see what happens in a pre-patterned cell with a variety of regions both with vanishing and non-zero $\rho = - \nabla \cdot \mathbf{P}$. We would expect to see    grain boundaries between different kinds of domain patterning. Finally, an important unexplored question is the nature of the domain wall between $\pi$-twisted domains.   We found that, given the other assumptions of our model such as a constant $\kappa$, our data were consistent with a domain wall energy density $f_{\pi} \propto h$, corresponding to a fixed width domain wall with elastic deformations spanning the entire cell thickness. It may be possible to have adjacent domains with the same twist chirality along the $z$ direction. In this case, one would expect a discontinuity in $\mathbf{P}$  orientation along the domain wall and possible uncompensated charge, leading to an even richer behavior.

\section*{Methods}
 
\subsection*{Materials and sample preparation}

We explore an $\mathrm{N}_\mathrm{F}$ material abbreviated DIO \cite{nishikawa2017fluid}, with a synthesis previously published in \cite{mandle2017rational}. On cooling from the isotropic (I) phase, the phase sequence of DIO is I-$174^{\circ}\mathrm{C}$-$\mathrm{N}$-$82^{\circ}\mathrm{C}$-$\mathrm{SmZ}_\mathrm{A}$-$66^{\circ}\mathrm{C}$-$\mathrm{N}_\mathrm{F}$-$34^{\circ}\mathrm{C}$-crystal, where N is a regular nematic and  $\mathrm{SmZ}_\mathrm{A}$ is an antiferroelectric smectic \cite{chen2023smectic}. The sandwich-type cells are bounded by two glass plates with layers of a photosensitive dye, Brilliant Yellow (BY), which shows maximum absorption in the range 400 nm to 550 nm. BY is dissolved in dimethylformamide (DMF) at a concentration 0.5 wt \%. The filtered BY-DMF solution is spin-coated onto the substrates at 3000 rpm for 30 seconds and baked for 30 minutes at $90^{\circ}\mathrm{C}$. The spin-coating and baking procedures are performed in a humidity-controlled environment with relative humidity fixed at 0.2.

To achieve apolar planar alignment, the BY-coated assembled cell is exposed to a light beam (light source EXFO X-Cite with a spectral range of 320 to 750 nm) with a linear polarizer for 10 minutes. This irradiation induces bidirectional molecular alignment perpendicular to the polarization axis of the normally incident light. The radial aster pattern of the dye molecules at the substrates is induced by irradiating the substrates through a plasmonic metamask with radial arrangements of nanoslits \cite{guo2016high,guo2016designs}. The bidirectional apolar anchoring is set by the same light beam that passes through the metamask and acquires local linear polarization orthogonal to the long axis of the nanoslit. The BY molecules realign perpendicularly to the local light polarization, thus the pattern of BY molecules replicates the pattern of nanoslits. To ensure that the surface patterns are the same on the top and bottom plates, these plates are assembled into an empty cell with a preset distance $h$ between them and irradiated by the same light beam. The cell is then filled with DIO in the N phase and then cooled down to the $\mathrm{N}_\mathrm{F}$ phase.

\subsection*{Numerical analysis}

The  estimates of the twist $\tau$  in Fig.~\ref{fig:Planar}j and in Fig.~\ref{fig:TD}c are performed by fitting the transmission data  to a Jones matrix model using $\tau $ as the fit parameter. The solid lines in these figures represent the fit for the value of $\tau$ given in the legend. The Jones matrix model is implemented in Wolfram Mathematica \cite{Mathematica}, and the relevant code and description (including all  parameters) are available in Ref.~\cite{sciencetwist2024}.

For the stripe width analysis in Fig.~\ref{fig:experiment}a, we minimize Eq.~\eqref{eq:stripeenergy}  with respect to $ \lambda_x$  using the ``FindMinimum'' routine in Wolfram Mathematica \cite{Mathematica}, keeping 500 terms in the summation, which does not appreciably change the result.
The expressions for $\lambda_{\mathrm{dw}}$ are varied as described in the Results section in order to get a reasonable match to the data in Fig.~\ref{fig:experiment}a.

 For the radial sector analysis [curves in Fig.~\ref{fig:experiment}(b)], the minimization of Eq.~\eqref{eq:fullenergy} with respect to $n_{\theta}$  is performed numerically, using the Nelder-Mead method implemented in the Scipy  Python library \cite{neldermead}. The associated integrals [the one explicitly in Eq.~\eqref{eq:fullenergy} and the one in the definition of $\Xi_{n_{\theta}}(z)$]  are evaluated using the method of quadrature, implemented via the Scipy library. The upper bound on the integration in Eq.~\eqref{eq:fullenergy} is taken to be 30 instead of infinite and we checked that lowering this bound did not change the value of the integral within the numerical error of the integrator.

 \bmhead{Acknowledgements}

We thank J. V. Selinger, L. Paik, D. Golovaty, and P. Sternberg for helpful discussions.
This work was supported by NSF  grant DMR-2341830 (O.D.L.\ and P.K.). After finalizing this work, we learned of a preprint by J. V. Selinger and L. Paik \cite{newselinger} that shows that, when confined to  a finite cell, the freely twisting ferroelectric nematic prefers to accommodate an integer number of twist pitches across the cell thickness.
\backmatter

\bibliography{sectors}% common bib file


\end{document}